\documentclass{article}

\PassOptionsToPackage{numbers, compress}{natbib}
\usepackage[final]{tackling_climate_workshop_style}

\usepackage[utf8]{inputenc} % allow utf-8 input
\usepackage[T1]{fontenc}    % use 8-bit T1 fonts
\usepackage{hyperref}       % hyperlinks
\usepackage{url}            % simple URL typesetting
\usepackage{booktabs}       % professional-quality tables
\usepackage{amsfonts}       % blackboard math symbols
\usepackage{nicefrac}       % compact symbols for 1/2, etc.
\usepackage{microtype}      % microtypography

\usepackage{nicefrac}       % compact symbols for 1/2, etc.
\usepackage{microtype}      % microtypography
\usepackage{xcolor}         % colors
\usepackage{graphicx}       % graphics

\usepackage{tabularx}
\usepackage{amsmath}
\usepackage{amssymb}
\usepackage{textcomp}
\usepackage{gensymb}
\usepackage{multibib}

\newcites{app}{Appendix References}

\graphicspath{{figures/}}

\definecolor{seagreen}{HTML}{2E8B57}
\definecolor{maroon}{HTML}{800000}
\definecolor{seablue}{HTML}{3585bf}

\title{Regional Ocean Forecasting with Hierarchical\\Graph Neural Networks}

\author{%
  Daniel Holmberg \\
  University of Helsinki \\
  \texttt{daniel.holmberg@helsinki.fi} \\
  \And
  Emanuela Clementi \\
  CMCC Foundation \\
  \texttt{emanuela.clementi@cmcc.it} \\
  \And
  Teemu Roos \\
  University of Helsinki \\
  \texttt{teemu.roos@helsinki.fi}
}

\begin{document}

\maketitle

\begin{abstract}
Accurate ocean forecasting systems are vital for understanding marine dynamics, which play a crucial role in environmental management and climate adaptation strategies. Traditional numerical solvers, while effective, are computationally expensive and time-consuming. Recent advancements in machine learning have revolutionized weather forecasting, offering fast and energy-efficient alternatives. Building on these advancements, we introduce SeaCast, a neural network designed for high-resolution, medium-range ocean forecasting. SeaCast employs a graph-based framework to effectively handle the complex geometry of ocean grids and integrates external forcing data tailored to the regional ocean context. Our approach is validated through experiments at a high spatial resolution using the operational numerical model of the Mediterranean Sea provided by the Copernicus Marine Service, along with both numerical and data-driven atmospheric forcings.
\end{abstract}

\section{Introduction}

Predicting sea dynamics is a formidable scientific challenge, driven by the need to understand and anticipate changes in ocean conditions that influence climate, weather forecasting, and maritime activities~\cite{le2019cmems}. While the need for improved ocean and coastal data is global, effective decision-making in sectors like shipping, resource management, and environmental monitoring often relies on regional and localized, high-resolution models able to deliver accurate forecasts~\cite{graham2018amm15, sun2019skrips, schiller2020bluelink, ciliberti2021blacksea, karna2021nemonordic, zhu2022scsofs}.

Recent advancements in machine learning-based weather prediction (MLWP) offer promising approaches to enhancing the predictability of marine conditions. By training on extensive historical datasets, these models can uncover complex patterns and dependencies, enabling predictions that are much faster and more energy-efficient than traditional numerical solvers. Several autoregressive ML models now rival or surpass the performance of the leading physics-based models used by meteorological organizations worldwide. Among the most notable advancements are the use of Transformers~\cite{bi2023panguweather, nguyen2024scaling, lang2024aifs}, neural operators~\cite{pathak2022fourcastnet, bonev2023spherical}, and graph neural networks (GNNs)~\cite{keisler2022forecasting, lam2023graphcast, oskarsson2023hilam}, each demonstrating significant potential in predicting atmospheric dynamics. 

However, the use of machine learning in ocean forecasting remains relatively nascent, and has primarily focused on climate emulation to date~\cite{ning2023graph, subel2024building, gray2024oceannet, guo2024orca, wang2024coupled}. This paper presents \emph{SeaCast}, an autoregressive model designed for high-resolution medium-range forecasting of regional oceans, extending the methodology from limited area weather forecasting using hierarchical GNNs~\cite{oskarsson2023hilam}. Our approach involves several key features allowing for accurate prediction of ocean states: 1) we adapt the graph creation, training, and evaluation processes to accommodate the irregular geometry of ocean grids; 2) the model integrates relevant atmospheric forcing near the sea surface; and 3) boundary forcing is applied to account for water in- and outflow, ensuring compatibility with the ocean at large. In our experiments, we train SeaCast using 35 years of reanalysis data followed by 2 years of analysis data of the Mediterranean Sea. We then compare the predictive performance on analysis fields and observations against an operational numerical forecaster, evaluating how well they forecast the state of the sea on a daily basis across 18 depth levels at a high spatial resolution of 1/24°.

\section{Method}

\paragraph{Problem Definition}
The forecasting problem is characterized by mapping a sequence of initial states $X^{-h:0} = (X^{-h}, ..., X^{0})$, where $h$ is the historical window, to a sequence of future states $X^{1:T} = (X^{1}, \dots, X^{T})$, where $T$ is the length of the forecast. Each state $X^t \in \mathbb{R}^{N \times d_x}$, contains $d_x$ variables at $N$ locations represented on a grid. Variables include both those at multiple vertical levels and single-level surface measures. In addition to this, forcing inputs $F^{1:T} = (F^{1}, ..., F^{T})$ encompassing known dynamic factors relevant to the forecasting problem, are also available.

\paragraph{Graph-based Neural Forecasting}
Initial states typically cover the two preceding time steps $X^{-1:0}$, enabling the capture of first-order dynamics~\cite{lam2023graphcast}. Concatenating a single-step forecast $\hat{X}^{t} = f(X^{t-2:t-1}, F^{t})$ to the initial state and repeating the process with $t$ incremented by one allows autoregressive forecasts $\hat{X}^{1:T}$ of arbitrary length $T$. An integral part of this framework is mapping from $N$ grid points onto a mesh graph $\mathcal{G}_M = (\mathcal{V}_M, \mathcal{E}_M)$ coarser than the simulation domain~\cite{pfaff2021learning}. The function $f$ is implemented as a sequence of GNN layers following an encode-process-decode architecture~\cite{sanchez2020learning} where: 1) grid inputs are encoded onto the mesh representation; 2) a number of GNN layers process this latent representation; 3) the processed data is mapped back onto the original grid. The mappings between grid and mesh nodes occur through bipartite grid-to-mesh $\mathcal{G}_{G2M}$ and mesh-to-grid $\mathcal{G}_{M2G}$ graphs. All node and edge updates are facilitated through GNN layers in the form of interaction networks~\cite{battaglia2016interactionnet} that map to a latent dimensionality $d_z$.

\paragraph{Hierarchical Graph}

The hierarchical mesh structure described in \cite{oskarsson2023hilam} consists of multiple graph levels $\mathcal{G}_1, \dots, \mathcal{G}_L$, where each level $\mathcal{G}_l = (\mathcal{V}_l, \mathcal{E}_l)$  gets progressively coarser. The first level connects directly to the grid, forming $\mathcal{G}_{G2M} = (\mathcal{V}_G \cup \mathcal{V}_1, \mathcal{E}_{G2M})$ and $\mathcal{G}_{M2G} = (\mathcal{V}_G \cup \mathcal{V}_1, \mathcal{E}_{M2G})$. A processing step on the mesh is defined as one sweep up and down through the hierarchy $\mathcal{G}_{l, l+1} = (\mathcal{V}_l \cup \mathcal{V}_{l+1}, \mathcal{E}_{l, l+1})$ where the graph sequences are $\mathcal{G}_{1,2}, ..., \mathcal{G}_{L-1,L}$ and $\mathcal{G}_{L,L-1}, ..., \mathcal{G}_{2,1}$, respectively.

\paragraph{Mesh Construction} Regional oceans can take very irregular shapes, calling for a customized mesh for the modeled geographical area. The foundation for our mesh is a quadrilateral construction used for graph-based limited area weather modeling~\cite{oskarsson2023hilam}. It is initialized by selecting only the nodes corresponding to the ocean surface grid. All nodes are connected with bidirectional edges to their neighbors horizontally, vertically and diagonally. Nodes on higher resolution levels are positioned in the center of a $3 \times 3$ square on the level below. Upward edges are created by connecting each node at level $l$ to the closest nodes at level $l+1$, and the downward edges mirror these. Edges crossing land areas with a threshold of 8 grid points are excluded, both for inter- and intra-level graphs. This procedure results in a mesh that conforms to the shape of the regional ocean.

\paragraph{Rollout Masking}

We want to ensure 1) that the learning task of the model is exclusively for grid nodes inside the regional ocean at each depth level, and 2) that the predictions are aligned with the influence of the global ocean. To address the first point we only propagate predictions part of the internal depth-wise ocean grid $\mathbb{G}$ in the autoregressive rollout. In response to the second point, boundary forcing is included at each time step by replacing predictions inside the boundary region $\mathbb{B}$ with the ground truth forecast $X_t$. We update the row for each node $v$ as:
\begin{equation}
    \hat{X}^{t}_{v,i} \leftarrow \left(\mathbb{I}_{\{v \in \mathbb{G}_{l(i)}\}} - \mathbb{I}_{\{v \in \mathbb{B}_{l(i)}\}}\right) \hat{X}^{t}_{v,i} + \mathbb{I}_{\{v \in \mathbb{B}_{l(i)}\}} X^{t}_{v,i}
\end{equation}  
Here, $\mathbb{I}_{\{\cdot\}}$ denotes the indicator function, equaling 1 if the specified condition is true and 0 otherwise. The set $\mathbb{G}_{l(i)}$ represents all oceanic grid nodes at the depth level $l(i)$ associated with feature $i$. The mapping function $l(i)$ assigns each feature to its corresponding depth level.

\paragraph{Training Objective}

The model is trained to minimize the mean squared error (MSE) between the predictions and the ground truth over a rolled-out sequence of states. The loss function we use is similar to what is commonly applied in MLWP, with the distinction that we account for the ocean grid structure at each depth level. The complete loss function is defined as:
\begin{equation}
    \mathcal{L} = \frac{1}{T_{\text{rollout}}} \sum_{t=1}^{T_{\text{rollout}}} \sum_{i=1}^{C} \sum_{l=1}^{L_i} \frac{1}{|\mathbb{G}_l|} \sum_{v \in \mathbb{G}_l} a_v \lambda_i  \left( \hat{X}^{t}_{v, i} - X^{t}_{v, i} \right)^2
\end{equation}
where $T_{\text{rollout}}$ is the number of steps in the rollout, $C$ is the number of feature channels in the tensor, $L_i$ is the number of depth levels for feature $i$, $\mathbb{G}_l$ is the set of ocean grid nodes at depth level $l$, $a_v$ is the latitude-longitude area of grid cell $v$ normalized to unit mean, and $\lambda_i$ is the inverse variance of time differences for variable $i$. Normalizing by the magnitude of the dynamics for each feature ensures that the model evaluates errors consistently across all vertical levels~\cite{keisler2022forecasting}.

\section{Experiments}

\paragraph{Dataset}
We evaluate our model using high-resolution numerical simulations of the Mediterranean Sea Physics (Med-PHY) provided by the Copernicus Marine Service (CMEMS). The dataset is configured to maintain causal separation between training, validation, and test sets. The training set includes reanalysis data~\cite{escudier2021reanalysis} from January 1987 to December 2021 and analysis data~\cite{clementi2023analysis} from January 2022 to April 2024. Validation is performed after each epoch using analysis data from May to June 2024. For testing, simulation states and analysis states from July to August 2024 are used, and the results are compared against the operational Med-PHY numerical forecasting system~\cite{clementi2023analysis} as well as a naive persistence model repeating the initial state over the whole forecasting period. We focus on the epipelagic zone of the sea by selecting every other available depth down to 200 meters, resulting in 18 vertical levels. The model forecasts seven different physical variables listed in Table~\ref{tab:variables}, three of which are single-level and four at multiple depths, resulting in 75 variables in total. All input variables are rescaled to zero mean and unit variance. Additionally, four static features—latitude, longitude, sea floor depth, and mean dynamic topography—are propagated, representing invariant characteristics at each grid node. This configuration leads to a total input grid dimension of $d_x=79$.

\paragraph{Atmospheric Forcing}

Atmospheric forcing play an important role in data-driven modeling of the ocean's response to atmospheric conditions~\cite{subel2024building}, especially in driving marine dynamics near the sea surface. We incorporate four key atmospheric variables: the 10-meter zonal and meridional wind components (\texttt{u10} and \texttt{v10}), the 2-meter temperature (\texttt{t2m}), mean sea level pressure (\texttt{msl}). The atmospheric data are sourced from daily mean aggregates of 6-hourly single-level ERA5 reanalysis data~\cite{hersbach2023era5}. For testing, we compare the 6-hourly daily means of the operational ECMWF numerical ensemble control forecast (ENS) and the new data-driven AIFS forecast~\cite{lang2024aifs}. All atmospheric forcing variables are bi-linearly interpolated from their native 1/4° resolution to the 1/24° resolution of the sea grid. Additionally, the sine and cosine of the day of year, normalized between 0 and 1, are included as forcing features to account for seasonal variations.
 
\paragraph{Boundary Forcing}

We define the boundary as the grid nodes west of longitude -5.2° to the edge the grid, effectively forcing the Straight of Gibraltar. Note that we use a boundary region that lies inside the propagated data grid, allowing us to use Mediterrenaen forecast data as boundary forcing. In an operational scenario a more principled approach could be to take the boundary forcing from a re-gridded global forecast. Current ocean forecasts at CMEMS are available 10 days in the future for the most part, following the length of HRES atmospheric forcing. However, we use the ENS/AIFS standard of 15-day forecasts. Hence we have to increase the length of the boundary forcing, and we do so by repeating the last forecast state in the boundary region five times at the end.

\paragraph{Model and Training}
We train SeaCast with 3 mesh levels and 4 processor layers with $d_z = 128$ hidden units, totaling 5.6M trainable parameters. Training initiates with a warm-up phase of five epochs, starting from a learning rate of $10^{-5}$ and incrementing epoch-wise to a base rate of $10^{-3}$. Following the warm-up, we employ a cosine decay schedule. For optimization, we use AdamW~\cite{loshchilov2019adamw}, configured with $\beta_1 = 0.9$, $\beta_2 = 0.95$, and $\lambda=0.1$. SeaCast is trained for 200 epochs using a batch size of 1. The number of rollout steps is progressively increased to 4 starting at 60\% of the total epochs. For 200 epochs this translates to updating the steps at epochs 120, 146, and 172 to 2, 3 and 4 rollout steps, respectively. The training took 2 days on 32 AMD MI250x GPUs.

\begin{figure}[t]
    \centering
    \includegraphics[width=\textwidth]{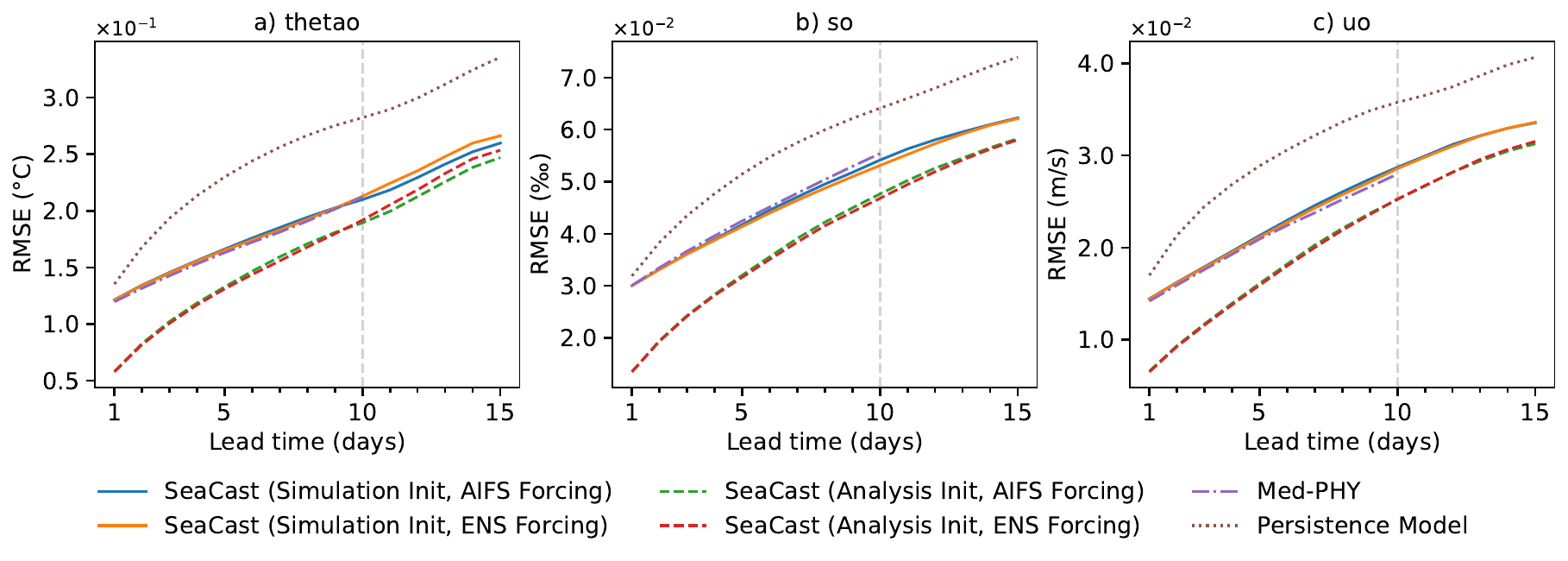}
    \caption{Depth-averaged RMSE for a) temperature, b) salinity, and c) zonal velocity.}
    \label{fig:thetao_so_uo_rmse}
\end{figure}

\paragraph{Results}
The forecasts are evaluated against analysis fields by calculating the Root Mean Squared Error (RMSE) for each lead time. The ML models outperform the persistence model across all lead times, and the ML models perform on par with Med-PHY for most variables, some of which are shown in Figure~\ref{fig:thetao_so_uo_rmse}. ML forecasts initialized with analysis data show significantly better performance. Currently, the Med-PHY operational system produces analysis fields to initialize the forecasts once a week on Tuesdays; however, it is expected that more frequent re-initialization using analysis fields could achieve the improved forecast skill shown here. More detailed results, including RMSE values across all depths and spatial error distributions, are provided in Appendix~\ref{app:additional_results}. The models are also compared to remotely sensed sea surface temperature (SST)~\cite{nardelli2013sstobs} and in situ measurements of temperature and salinity. With respect to Med-PHY, the ML models compare favorably for SST forecasting by a good margin, and against in situ observations all models show similar results.

\begin{figure}[t]
    \centering
    \includegraphics[width=\textwidth]{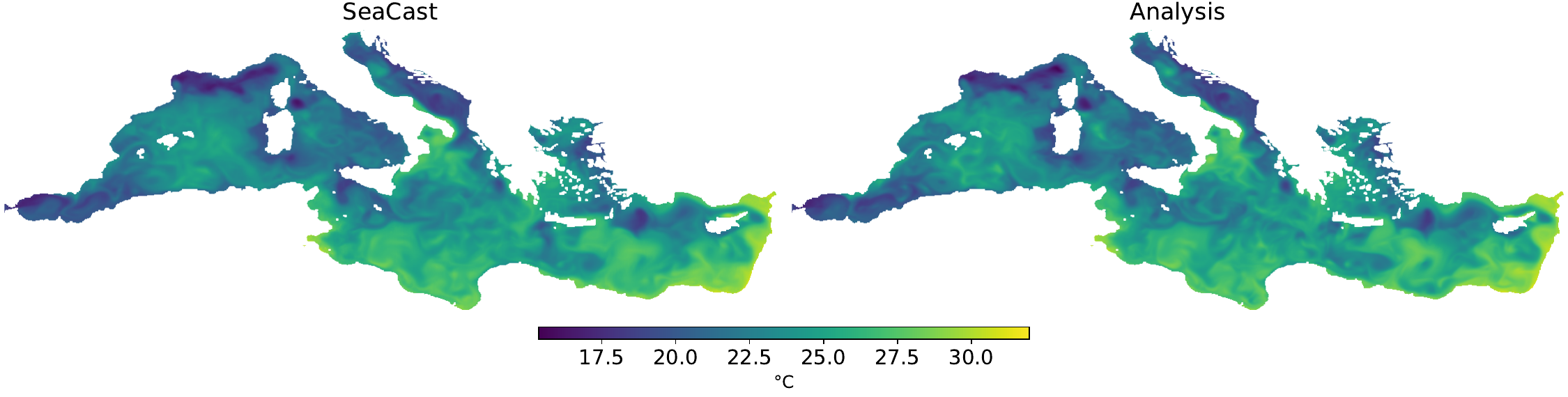}
    \caption{Comparison of the SeaCast (AIFS forcing) 10-day lead temperature forecast at a depth of 22.7 m, initialized from simulation on August 1st, 2024, against the corresponding analysis field.}
    \label{fig:rollout_thetao_23}
\end{figure}

\section{Conclusion}

We introduced SeaCast, an autoregressive machine learning model designed for high-resolution, medium-range ocean forecasting, capable of handling complex sea surface geometries and incorporating relevant atmospheric forcings. Our results show that SeaCast achieves comparable performance to the operational Med-PHY model for the Mediterranean Sea, while operating at a fraction of the computational cost. Future enhancements could involve using higher temporal resolution training data to expand the sample size and provide more granular predictions. Applying this approach to other seas or incorporating different oceanographic variables is another promising direction to go.

\begin{ack}

This work was financially supported by the Research Council of Finland under the ICT 2023: Frontier AI Technologies program (Grant No. 345635). Computing resources were provided by the LUMI supercomputer, owned by the EuroHPC Joint Undertaking and hosted by CSC–IT Center for Science.

%This study has been conducted using open data from Copernicus Marine Service, Copernicus Climate Change Service, and European Centre for Medium-Range Weather Forecasts.

%This study has been conducted using E.U. Copernicus Marine Service information: \texttt{doi:10.25423/cmcc/medsea\_analysisforecast\_phy\_006\_013\_eas8}, \texttt{doi:10.25423/cmcc/medsea\_multiyear\_phy\_006\_004\_e3r1}, \texttt{doi:10.48670/moi-00044}, \texttt{doi:10.48670/moi-00172}; European Centre for Medium-Range Weather Forecasts information: \texttt{doi:10.21957/open-data}, and Copernicus Climate Change Service information: \texttt{doi:10.24381/cds.f17050d7}.

\end{ack}

\bibliographystyle{unsrt}  
\bibliography{references}

\begin{thebibliography}{10}

\bibitem{liubartseva2023oilspill}
S.~Liubartseva, G.~Coppini, G.~Verdiani, T.~Mungari, F.~Ronco, M.~Pinto, G.~Pastore, and R.~Lecci.
\newblock Modeling chronic oil pollution from ships.
\newblock {\em Marine Pollution Bulletin}, 2023.

\bibitem{mannarini2024visir}
Gianandrea Mannarini, Mario~Leonardo Salinas, Lorenzo Carelli, Nicola Petacco, and Josip Orovi{\'c}.
\newblock {VISIR-2: Ship weather routing in Python}.
\newblock {\em Geoscientific Model Development}, 17(10):4355--4382, 2024.

\bibitem{coppini2023forecast}
Giovanni Coppini, Emanuela Clementi, Gianpiero Cossarini, Stefano Salon, Gerasimos Korres, Michalis Ravdas, Rita Lecci, Jenny Pistoia, Anna~Chiara Goglio, Massimiliano Drudi, et~al.
\newblock {The Mediterranean forecasting system—Part 1: Evolution and performance}.
\newblock {\em Ocean Science}, 19(5):1483--1516, 2023.

\bibitem{ben2024rise}
Zied Ben~Bouall{\`e}gue, Mariana~CA Clare, Linus Magnusson, Estibaliz Gascon, Michael Maier-Gerber, Martin Janou{\v{s}}ek, Mark Rodwell, Florian Pinault, Jesper~S Dramsch, Simon~TK Lang, et~al.
\newblock The rise of data-driven weather forecasting: A first statistical assessment of machine learning--based weather forecasts in an operational-like context.
\newblock {\em Bulletin of the American Meteorological Society}, 105(6):E864--E883, 2024.

\bibitem{ecmwf2024plans}
ECMWF.
\newblock Plans for high resolution forecast ({HRES}) and ensemble forecast ({ENS}).
\newblock {\em In focus}, March 2024.
\newblock URL \url{https://www.ecmwf.int/en/about/media-centre/focus/2024/plans-high-resolution-forecast-hres-and-ensemble-forecast-ens}.

\bibitem{sein2015rom}
Dmitry~V Sein, Uwe Mikolajewicz, Matthias Gr{\"o}ger, Irina Fast, William Cabos, Joaquim~G Pinto, Stefan Hagemann, Tido Semmler, Alfredo Izquierdo, and Daniela Jacob.
\newblock Regionally coupled atmosphere-ocean-sea ice-marine biogeochemistry model {ROM}: 1. {Description} and validation.
\newblock {\em Journal of Advances in Modeling Earth Systems}, 7(1):268--304, 2015.

\bibitem{nguyen2023climax}
Tung Nguyen, Johannes Brandstetter, Ashish Kapoor, Jayesh~K Gupta, and Aditya Grover.
\newblock {ClimaX}: A foundation model for weather and climate.
\newblock In {\em International Conference on Machine Learning}, 2023.

\bibitem{bodnar2024aurora}
Cristian Bodnar, Wessel~P Bruinsma, Ana Lucic, Megan Stanley, Johannes Brandstetter, Patrick Garvan, Maik Riechert, Jonathan Weyn, Haiyu Dong, Anna Vaughan, et~al.
\newblock Aurora: A foundation model of the atmosphere.
\newblock {\em arXiv preprint arXiv:2405.13063}, 2024.

\bibitem{price2023gencast}
Ilan Price, Alvaro Sanchez-Gonzalez, Ferran Alet, Timo Ewalds, Andrew El-Kadi, Jacklynn Stott, Shakir Mohamed, Peter Battaglia, Remi Lam, and Matthew Willson.
\newblock {GenCast}: Diffusion-based ensemble forecasting for medium-range weather.
\newblock {\em arXiv preprint arXiv:2312.15796}, 2023.

\bibitem{oskarsson2024probabilistic}
Joel Oskarsson, Tomas Landelius, Marc~Peter Deisenroth, and Fredrik Lindsten.
\newblock Probabilistic weather forecasting with hierarchical graph neural networks.
\newblock {\em arXiv preprint arXiv:2406.04759}, 2024.

\bibitem{madec2017nemo}
Gurvan Madec, Romain Bourdall{\'e}-Badie, Pierre-Antoine Bouttier, Cl{\'e}ment Bricaud, Diego Bruciaferri, Daley Calvert, J{\'e}r{\^o}me Chanut, Emanuela Clementi, Andrew Coward, Damiano Delrosso, et~al.
\newblock Nemo ocean engine.
\newblock {\em Scientific Notes of Climate Modelling Center}, 27, 2017.

\bibitem{dobricic2008oceanvar}
Srdjan Dobricic and Nadia Pinardi.
\newblock An oceanographic three-dimensional variational data assimilation scheme.
\newblock {\em Ocean modelling}, 22(3-4):89--105, 2008.

\bibitem{weatherall2015bathymetry}
Pauline Weatherall, Karen~M Marks, Martin Jakobsson, Thierry Schmitt, Shin Tani, Jan~Erik Arndt, Marzia Rovere, Dale Chayes, Vicki Ferrini, and Rochelle Wigley.
\newblock A new digital bathymetric model of the world's oceans.
\newblock {\em Earth and space Science}, 2(8):331--345, 2015.

\bibitem{ramachandran2017swish}
Prajit Ramachandran, Barret Zoph, and Quoc~V Le.
\newblock Searching for activation functions.
\newblock {\em arXiv preprint arXiv:1710.05941}, 2017.

\bibitem{ba2016layernorm}
Jimmy~Lei Ba, Jamie~Ryan Kiros, and Geoffrey~E. Hinton.
\newblock Layer normalization.
\newblock {\em arXiv preprint arXiv:1607.06450}, 2016.

\bibitem{cmems2024sst}
E.U. Copernicus~Marine Service.
\newblock Mediterranean sea high resolution and ultra high resolution sea surface temperature analysis.
\newblock {\em Marine Data Store (MDS)}, 2024.
\newblock DOI \href{https://doi.org/10.48670/moi-00172}{\texttt{10.48670/moi-00172}}.

\bibitem{cmems2024insitu}
E.U. Copernicus~Marine Service.
\newblock Mediterranean sea- in-situ near real time observations.
\newblock {\em Marine Data Store (MDS)}, 2024.
\newblock DOI \href{https://doi.org/10.48670/moi-00044}{\texttt{10.48670/moi-00044}}.

\end{thebibliography}


\begin{thebibliography}{10}

\bibitem{le2019cmems}
Pierre~Yves Le~Traon, Antonio Reppucci, Enrique Alvarez~Fanjul, Lotfi Aouf, Arno Behrens, Maria Belmonte, Abderrahim Bentamy, Laurent Bertino, Vittorio~Ernesto Brando, Matilde~Brandt Kreiner, et~al.
\newblock From observation to information and users: The {Copernicus Marine Service} perspective.
\newblock {\em Frontiers in Marine Science}, 6:234, 2019.

\bibitem{graham2018amm15}
Jennifer~A Graham, Enda O'Dea, Jason Holt, Jeff Polton, Helene~T Hewitt, Rachel Furner, Karen Guihou, Ashley Brereton, Alex Arnold, Sarah Wakelin, et~al.
\newblock {AMM15: A new high-resolution NEMO configuration for operational simulation of the European north-west shelf}.
\newblock {\em Geoscientific Model Development}, 11(2):681--696, 2018.

\bibitem{sun2019skrips}
Rui Sun, Aneesh~C Subramanian, Arthur~J Miller, Matthew~R Mazloff, Ibrahim Hoteit, and Bruce~D Cornuelle.
\newblock {SKRIPS} v1.0: A regional coupled ocean--atmosphere modeling framework ({MITgcm--WRF}) using {ESMF/NUOPC}, description and preliminary results for the {Red Sea}.
\newblock {\em Geoscientific Model Development}, 12(10):4221--4244, 2019.

\bibitem{schiller2020bluelink}
Andreas Schiller, Gary~B Brassington, Peter Oke, Madeleine Cahill, Prasanth Divakaran, Mikhail Entel, Justin Freeman, David Griffin, Mike Herzfeld, Ron Hoeke, et~al.
\newblock Bluelink ocean forecasting {Australia}: 15 years of operational ocean service delivery with societal, economic and environmental benefits.
\newblock {\em Journal of Operational Oceanography}, 13(1):1--18, 2020.

\bibitem{ciliberti2021blacksea}
Stefania~A Ciliberti, Marilaure Gr{\'e}goire, Joanna Staneva, Atanas Palazov, Giovanni Coppini, Rita Lecci, Elisaveta Peneva, Marius Matreata, Veselka Marinova, Simona Masina, et~al.
\newblock Monitoring and forecasting the ocean state and biogeochemical processes in the {Black Sea}: Recent developments in the {Copernicus Marine Service}.
\newblock {\em Journal of Marine Science and Engineering}, 9(10):1146, 2021.

\bibitem{karna2021nemonordic}
Tuomas K{\"a}rn{\"a}, Patrik Ljungemyr, Saeed Falahat, Ida Ringgaard, Lars Axell, Vasily Korabel, Jens Murawski, Ilja Maljutenko, Anja Lindenthal, Simon Jandt-Scheelke, et~al.
\newblock {Nemo-Nordic 2.0}: Operational marine forecast model for the {Baltic Sea}.
\newblock {\em Geoscientific Model Development}, 14(9):5731--5749, 2021.

\bibitem{zhu2022scsofs}
Xueming Zhu, Ziqing Zu, Shihe Ren, Miaoyin Zhang, Yunfei Zhang, Hui Wang, and Ang Li.
\newblock Improvements in the regional {South China Sea} operational oceanography forecasting system ({SCSOFSv2}).
\newblock {\em Geoscientific Model Development}, 15(3):995--1015, 2022.

\bibitem{bi2023panguweather}
Kaifeng Bi, Lingxi Xie, Hengheng Zhang, Xin Chen, Xiaotao Gu, and Qi~Tian.
\newblock Accurate medium-range global weather forecasting with {3D} neural networks.
\newblock {\em Nature}, 619(7970):533--538, 2023.

\bibitem{nguyen2024scaling}
Tung Nguyen, Rohan Shah, Hritik Bansal, Troy Arcomano, Sandeep Madireddy, Romit Maulik, Veerabhadra Kotamarthi, Ian Foster, and Aditya Grover.
\newblock Scaling transformers for skillful and reliable medium-range weather forecasting.
\newblock In {\em ICLR 2024 Workshop on Tackling Climate Change with Machine Learning}, 2024.

\bibitem{lang2024aifs}
Simon Lang, Mihai Alexe, Matthew Chantry, Jesper Dramsch, Florian Pinault, Baudouin Raoult, Mariana~CA Clare, Christian Lessig, Michael Maier-Gerber, Linus Magnusson, et~al.
\newblock {AIFS - ECMWF}'s data-driven forecasting system.
\newblock {\em arXiv preprint arXiv:2406.01465}, 2024.

\bibitem{pathak2022fourcastnet}
Jaideep Pathak, Shashank Subramanian, Peter Harrington, Sanjeev Raja, Ashesh Chattopadhyay, Morteza Mardani, Thorsten Kurth, David Hall, Zongyi Li, Kamyar Azizzadenesheli, et~al.
\newblock {FourCastNet}: A global data-driven high-resolution weather model using adaptive {Fourier} neural operators.
\newblock {\em arXiv preprint arXiv:2202.11214}, 2022.

\bibitem{bonev2023spherical}
Boris Bonev, Thorsten Kurth, Christian Hundt, Jaideep Pathak, Maximilian Baust, Karthik Kashinath, and Anima Anandkumar.
\newblock Spherical {Fourier} neural operators: Learning stable dynamics on the sphere.
\newblock In {\em International Conference on Machine Learning}, pages 2806--2823. PMLR, 2023.

\bibitem{keisler2022forecasting}
Ryan Keisler.
\newblock Forecasting global weather with graph neural networks.
\newblock {\em arXiv preprint arXiv:2202.07575}, 2022.

\bibitem{lam2023graphcast}
Remi Lam, Alvaro Sanchez-Gonzalez, Matthew Willson, Peter Wirnsberger, Meire Fortunato, Ferran Alet, Suman Ravuri, Timo Ewalds, Zach Eaton-Rosen, Weihua Hu, Alexander Merose, Stephan Hoyer, George Holland, Oriol Vinyals, Jacklynn Stott, Alexander Pritzel, Shakir Mohamed, and Peter Battaglia.
\newblock Learning skillful medium-range global weather forecasting.
\newblock {\em Science}, 382(6677):1416--1421, 2023.

\bibitem{oskarsson2023hilam}
Joel Oskarsson, Tomas Landelius, and Fredrik Lindsten.
\newblock Graph-based neural weather prediction for limited area modeling.
\newblock In {\em NeurIPS 2023 Workshop on Tackling Climate Change with Machine Learning}, 2023.

\bibitem{ning2023graph}
Ding Ning, Varvara Vetrova, and Karin~R Bryan.
\newblock Graph-based deep learning for sea surface temperature forecasts.
\newblock In {\em ICLR 2023 Workshop on Tackling Climate Change with Machine Learning}, 2023.

\bibitem{subel2024building}
Adam Subel and Laure Zanna.
\newblock Building ocean climate emulators.
\newblock In {\em ICLR 2024 Workshop on Tackling Climate Change with Machine Learning}, 2024.

\bibitem{gray2024oceannet}
Michael~A Gray, Ashesh Chattopadhyay, Tianning Wu, Anna Lowe, and Ruoying He.
\newblock Long-term prediction of the {Gulf Stream} meander using {OceanNet}: A principled neural operator-based digital twin.
\newblock {\em EGUsphere}, 2024:1--23, 2024.

\bibitem{guo2024orca}
Zijie Guo, Pumeng Lyu, Fenghua Ling, Jing-Jia Luo, Niklas Boers, Wanli Ouyang, and Lei Bai.
\newblock {ORCA}: A global ocean emulator for multi-year to decadal predictions.
\newblock {\em arXiv preprint arXiv:2405.15412}, 2024.

\bibitem{wang2024coupled}
Chenggong Wang, Michael~S. Pritchard, Noah Brenowitz, Yair Cohen, Boris Bonev, Thorsten Kurth, Dale Durran, and Jaideep Pathak.
\newblock Coupled ocean-atmosphere dynamics in a machine learning earth system model.
\newblock {\em arXiv preprint arXiv: 2406.08632}, 2024.

\bibitem{pfaff2021learning}
Tobias Pfaff, Meire Fortunato, Alvaro Sanchez-Gonzalez, and Peter Battaglia.
\newblock Learning mesh-based simulation with graph networks.
\newblock In {\em International Conference on Learning Representations}, 2021.

\bibitem{sanchez2020learning}
Alvaro Sanchez-Gonzalez, Jonathan Godwin, Tobias Pfaff, Rex Ying, Jure Leskovec, and Peter Battaglia.
\newblock Learning to simulate complex physics with graph networks.
\newblock In {\em International Conference on Machine Learning}, pages 8459--8468. PMLR, 2020.

\bibitem{battaglia2016interactionnet}
Peter Battaglia, Razvan Pascanu, Matthew Lai, Danilo Jimenez~Rezende, et~al.
\newblock Interaction networks for learning about objects, relations and physics.
\newblock {\em Advances in neural information processing systems}, 29, 2016.

\bibitem{escudier2021reanalysis}
Romain Escudier, Emanuela Clementi, Andrea Cipollone, Jenny Pistoia, Massimiliano Drudi, Alessandro Grandi, Vladislav Lyubartsev, Rita Lecci, Ali Aydogdu, Damiano Delrosso, et~al.
\newblock A high resolution reanalysis for the {Mediterranean Sea}.
\newblock {\em Frontiers in Earth Science}, 9:702285, 2021.

\bibitem{clementi2023analysis}
E.~Clementi, M.~Drudi, A.~Aydogdu, A.~Moulin, A.~Grandi, A.~Mariani, A.~C. Goglio, J.~Pistoia, P.~Miraglio, R.~Lecci, F.~Palermo, G.~Coppini, S.~Masina, and N.~Pinardi.
\newblock {Mediterranean Sea} physical analysis and forecast (version 1).
\newblock {\em Copernicus Monitoring Environment Marine Service ({CMEMS})}, 2023.

\bibitem{hersbach2023era5}
H.~Hersbach, B.~Bell, P.~Berrisford, G.~Biavati, A.~Horányi, J.~Muñoz~Sabater, J.~Nicolas, C.~Peubey, R.~Radu, I.~Rozum, D.~Schepers, A.~Simmons, C.~Soci, D.~Dee, and J-N. Thépaut.
\newblock {ERA5} monthly averaged data on single levels from 1940 to present.
\newblock {\em Copernicus Climate Change Service (C3S) Climate Data Store (CDS)}, 2023.

\bibitem{loshchilov2019adamw}
Ilya Loshchilov and Frank Hutter.
\newblock Decoupled weight decay regularization.
\newblock In {\em International Conference on Learning Representations}, 2019.

\bibitem{nardelli2013sstobs}
B~Buongiorno Nardelli, C~Tronconi, A~Pisano, and R~Santoleri.
\newblock {High and Ultra-High resolution processing of satellite Sea Surface Temperature data over Southern European Seas in the framework of MyOcean project}.
\newblock {\em Remote Sensing of Environment}, 129:1--16, 2013.

\end{thebibliography}

\newpage

\appendix

\section{Code and Data Availability}

The source code for SeaCast is available at \url{https://github.com/deinal/seacast}. This work used publicly available data from Copernicus Marine Service and the European Centre for Medium Range Forecasting for training and evaluation. Predictions and test data based on forecasts from July to August 2024 are stored at \url{https://doi.org/10.5281/zenodo.13894915}.

\section{Climate Impact}

\paragraph{Energy Footprint} Traditional numerical weather prediction systems rely on large-scale computing clusters, consuming large amounts of energy to operate. Learned forecasting systems like the one presented here offer a more energy-efficient alternative, as only the training phase requires substantial compute but inference runs very cheaply.

\paragraph{Regional Ocean Forecasting} Global and local models play crucial roles in understanding and responding to climate change. While global models can inform global decision-making, local models enable detailed studies of specific regions, empowering local communities and national agencies to tailor their climate adaptation strategies. Accurate and localized forecasts assist in managing marine resources, preparing for extreme weather events, and mitigating climate impacts on coastal and marine ecosystems, such as monitoring oil spill events~\citeapp{liubartseva2023oilspill}, tracking optimal ship routing~\citeapp{mannarini2024visir}, and many other use cases\footnote{\url{https://marine.copernicus.eu/services/use-cases}}.

\section{Future Work}

\paragraph{Improvements to the Reanalysis} The Mediterranean Sea reanalysis currently lacks some features present in the operational Med-PHY system, such as open boundaries at the Dardanelles Strait, tidal inputs, and coupling with a wave model~\citeapp{coppini2023forecast}, which could be helpful for training ML models. Furthermore, storing higher temporal resolution outputs, such as 6-hourly sea states, from the reanalysis would significantly increase the amount of available training data, and allow the models to capture changes in the sea that follow the diurnal cycle.

\paragraph{Evaluation} Our current evaluation approach involves comparing the model outputs against analysis fields of the Mediterranean Sea and a limited period of observations. While this provides a baseline for assessing model performance, it could be improved by evaluating on a much longer timespan to better assess the model's ability to capture seasonal and interannual variability. Also, comparing with up-to-date reanalysis data could offer a more accurate representation of the forecast skill. The model would benefit from a statistically more exhaustive evaluation~\citeapp{ben2024rise}, including assessing how well the model can forecast extreme events.

\paragraph{Forecast Length} With the recent updates in the Integrated Forecasting System to Cycle 48r1, the medium-range ensemble forecasts have been extended from 10 days to 15 days, replacing the previous 10-day high-resolution forecast~\citeapp{ecmwf2024plans}. This update has been noted when evaluating our model. However, we could not compare the last five days with the Mediterranean operational forecasting system as the open CMEMS products still obey the 10-day standard. However, this is expected to change some time in the future.

\paragraph{Coupled Models} Coupling has been shown to improve the accuracy of numerical ocean models by integrating multiple components, such as ocean physics, biogeochemistry, atmospheric models, and sea ice simulations~\citeapp{sein2015rom}. This approach effectively captures the complex feedback mechanisms between systems, leading to improved predictive fidelity. This concept could be extended to neural ocean forecasting, where training on both ocean physics and biogeochemistry simulations, for example, could yield more accurate and holistic predictions. However, this depends on the availability of historical datasets and the specific requirements of the region being studied.

\paragraph{Foundation Models} Foundation models has been applied for atmospheric forecasting~\citeapp{nguyen2023climax, bodnar2024aurora} and offer a promising avenue for enhancing ocean prediction. These models can improve forecast accuracy by learning general-purpose representations of atmospheric dynamics through training on large, diverse datasets, including simulations from various weather and climate models. Applying this concept to ocean forecasting could yield similar benefits, enabling models to better generalize across different oceanographic contexts by learning from different data sources and adapt to new prediction tasks more effectively.

\paragraph{Probabilistic Forecasting} While our current approach is based on deterministic forecasts similar to the operational numerical forecaster, the incorporation of probabilistic forecasting and ensemble methods, as seen in recent machine learning developments~\citeapp{price2023gencast, oskarsson2024probabilistic}, could provide a deeper understanding of forecast uncertainty and produce more physically realistic ensemble members.

\section{Data Details}
\label{app:data_details}

\paragraph{Dataset} The Mediterranean Sea physics analysis and reanalysis systems leverage the coupled hydrodynamic-wave model Nucleus for European Modelling of the Ocean (NEMO)~\citeapp{madec2017nemo} and the variational data assimilation scheme OceanVar~\citeapp{dobricic2008oceanvar} to simulate ocean dynamics and integrate in situ measurements. The simulations are conducted at a high horizontal grid resolution of 1/24° (approximately 4 km), utilizing a total of 141 vertical levels with uneven spacing. By choosing every other depth until 200 m the list of selected depths becomes: 1.02, 5.46, 10.5, 16.3, 22.7, 29.9, 37.9, 46.7, 56.3, 66.9, 78.6, 91.2, 105, 120, 136, 153, 172, and 192 meters. The topography is based on an interpolation of the General Bathymetric Chart of the Oceans (GEBCO) 30 arcsecond grid~\citeapp{weatherall2015bathymetry}. The test data consists of forecasts issued from July 24th to August 4th, 2024. The last date used for evaluation is hence August 19th, 2024, and the total number of days in the set is 27.

\begin{table}[t]
\centering
\caption{Summary of all variables, static fields, and forcing features in the sea physics dataset.}
\begin{tabular}{llll}
\toprule
 & Abbreviation & Unit & Vertical Level \\
\midrule
Variables & & & \\
\midrule
Eastward sea water velocity & \texttt{uo} & m/s & 18 depths \\
Northward sea water velocity & \texttt{vo} & m/s & 18 depths \\
Ocean mixed layer thickness & \texttt{mlotst} & m & Sea surface \\
Sea water salinity & \texttt{so} & \textperthousand & 18 depths \\
Sea surface height above geoid & \texttt{zos} & m & Sea surface \\
Sea water potential temperature & \texttt{thetao} & \textdegree C & 18 depths \\
Sea water potential temperature at the sea floor & \texttt{bottomT} & \textdegree C & Sea floor \\
\midrule
Static fields & & & \\
\midrule
Sea floor depth below geoid & \texttt{deptho} & m & Sea floor \\
Mean dynamic topography & \texttt{mdt} & m & Sea surface \\
Latitude & \texttt{lat} & \textdegree & - \\
Longitude & \texttt{lon} & \textdegree & - \\
\midrule
Forcing & & & \\
\midrule
10m u-component of wind & \texttt{u10} & m/s & 10 m above surface \\
10m v-component of wind & \texttt{v10} & m/s & 10 m above surface \\
2m temperature & \texttt{t2m} & \textdegree C & 2 m above surface \\
Mean sea level pressure & \texttt{msl} & Pa & Sea surface \\
Sine of time of year & \texttt{sin\_toy} & - & - \\
Cosine of time of year & \texttt{cos\_toy} & - & - \\
\bottomrule
\end{tabular}
\label{tab:variables}
\end{table}

\begin{figure}[h]
    \centering
    \includegraphics[width=\textwidth]{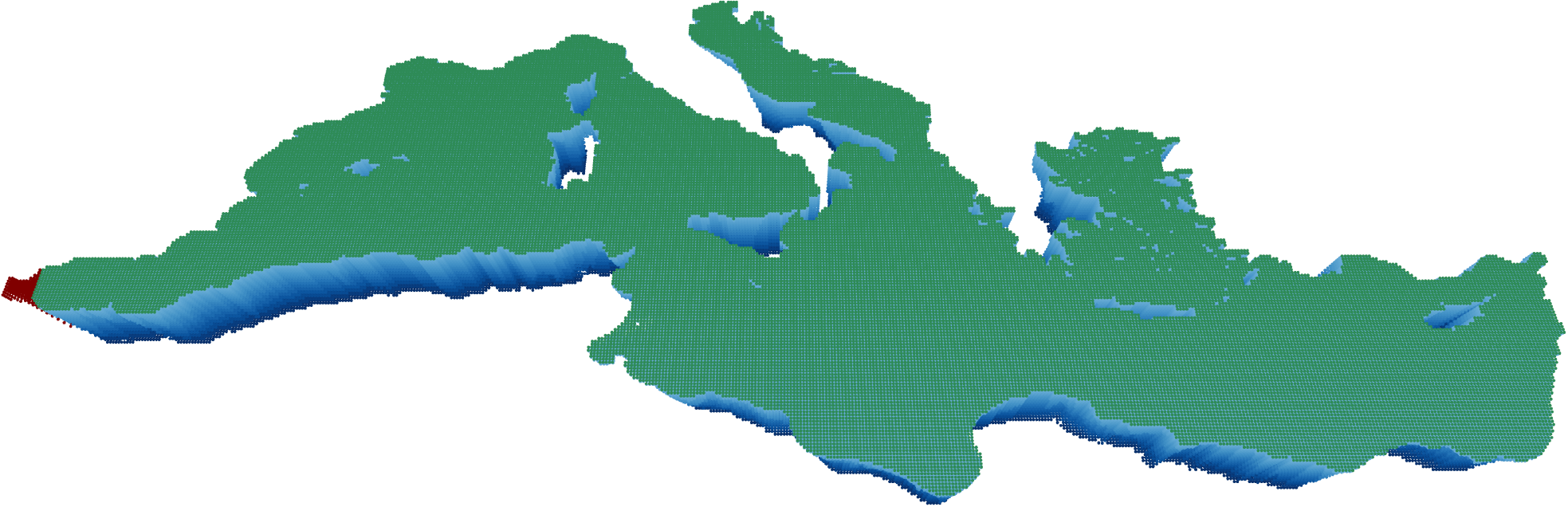}
    \caption{Illustration of the Mediterranean Sea at all 18 depth levels. The surface layer is colored \textcolor{seagreen}{seagreen} and the boundary forcing region in the Strait of Gibraltar is colored \textcolor{maroon}{maroon}. The color of the interior sea gets darker \textcolor{seablue}{blue} at increased depth. The visualized depth relative to the height and width is not to scale.}
    \label{fig:grid}
\end{figure}

\paragraph{Data Grid} The minimum rectangle in which the Mediterranean Sea fits at the current resolution is 371 by 1013 grid points. This is a total of 375 823 points, whereas the actual sea surface of the Mediterranean only has $N$ = 144 990 points, or 45\% of the total number. Only this subset of grid points has to be processed by the GNN. The complete data grid is shown in Figure~\ref{fig:grid}. The forcing region is highlighted to the left in that plot, where the Strait of Gribraltar connects the Mediterranean Sea with the Atlantic Ocean. Single level features, forcing features, and static fields all cover the surface area, whereas variables at multiple depths have values on all vertical levels shown in the plot.

\section{Model Details}
\label{app:model_details}

\begin{figure}[h]
    \centering
    \includegraphics[width=\textwidth]{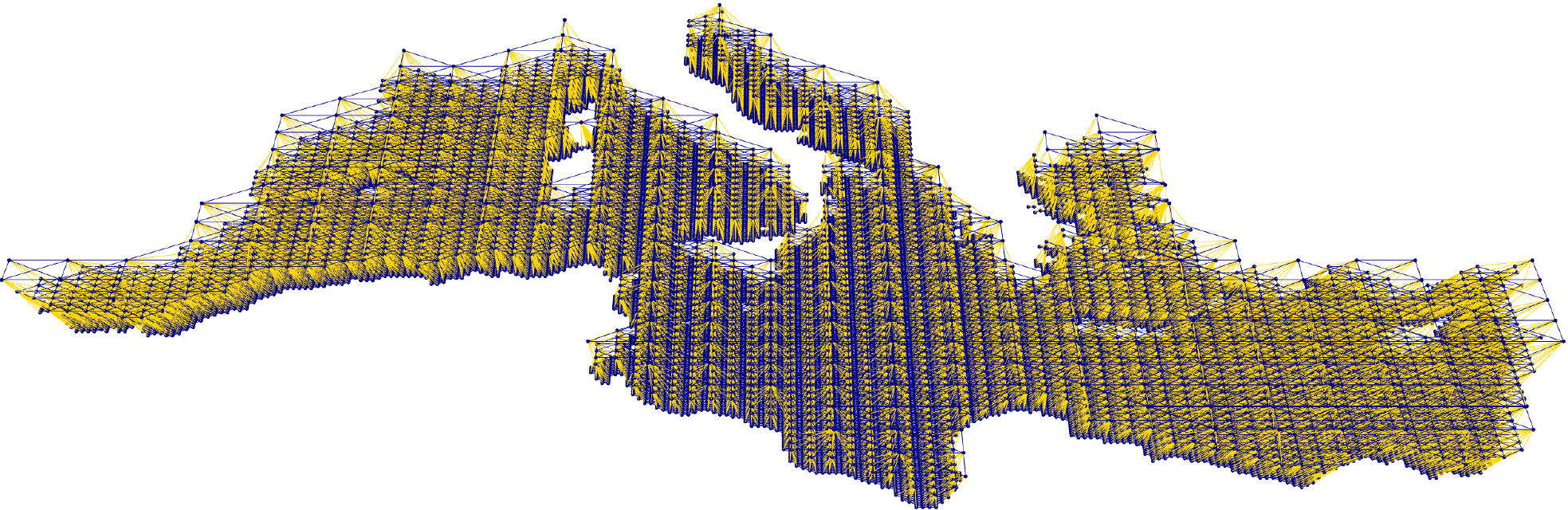}
    \caption{Ocean variables are encoded onto a hierarchical mesh of the Mediterranean Sea shown here. Each layer has a different resolution allowing for interactions at different scales between observables.}
    \label{fig:hi_graph}
\end{figure}

\paragraph{Network Structure} The 3-layer hierarchical graph used for SeaCast is shown in Figure~\ref{fig:hi_graph}, and the statistics defining this graph is shown in Table~\ref{tab:graph}. Multilayer perceptrons (MLP) within GNN layers consists of a single hidden layer using the Swish activation function~\citeapp{ramachandran2017swish}, followed by layer normalization~\citeapp{ba2016layernorm}.

\paragraph{Atmospheric Forcing} The atmospheric forcing used in our model is windowed over three consecutive time steps. This means that each forcing input $F_t$ includes data from times $t - 1$, $t$, and $t + 1$. For the past atmospheric forcing inputs at time steps -1 and 0, the test set utilizes data from the daily aggregates of one-day-ahead atmospheric forecasts issued on the preceding days. However, these inputs could alternatively come from a product that provides a more accurate representation of atmospheric conditions for recent days.

\begin{table}[h]
\centering
\caption{Number of nodes and edges in the Mediterranean graph.}
\begin{tabular}{@{}lll@{}}
\toprule
Graph                   & Nodes     & Edges \\
\midrule
$\mathcal{G}_0$         & 22677     & 174007 \\
$\mathcal{G}_{0, 1} / \mathcal{G}_{1,0}$ & -        & 22677  \\
$\mathcal{G}_1$         & 2515      & 18206    \\
$\mathcal{G}_{1, 2} / \mathcal{G}_{2, 1}$ & -        & 2515   \\
$\mathcal{G}_2$         & 272       & 1610 \\
\midrule
Mesh             & 25464    & 219015  \\
\midrule
$\mathcal{G}_{G2M}$     & -       & 542271 \\
$\mathcal{G}_{M2G}$     & -       & 579960 \\
\midrule
Grid                    & 144990   & - \\
\bottomrule
\end{tabular}
\label{tab:graph}
\end{table}

\paragraph{Computational Complexity}

Training the SeaCast model took 2 days on 32 AMD MI250x GPUs. SeaCast can then produce a complete 15-day forecast in 11.2 seconds on a single GPU, which is roughly equivalent to 0.75 seconds per timestep. The SeaCast forecast includes 18 depth levels, and outputs predictions at a daily temporal resolution. In contrast, the computational time required for the Med-PHY system is approximately 135 minutes to run a bulletin, consisting of a 1-day simulation followed by a 10-day forecast, using 413 CPU cores. This includes the time for generating and writing outputs for all 141 vertical levels, and running a coupled wave model. The model timestep is 240 seconds for reanalysis and 120 seconds for analysis/forecast, with reanalysis outputs provided as daily mean values and forecast outputs available at both 1-hour and daily mean frequencies. All systems produce outputs at the same 1/24° spatial resolution.

\section{Additional Results}
\label{app:additional_results}

In the results below, SeaCast with simulation initialization and AIFS forcing is denoted as SeaCast-AIFS, and SeaCast with simulation initialization and ENS forcing is denoted as SeaCast-ENS.

\subsection{Comparison to Satellite SST}

\begin{figure}[h]
    \centering
    \includegraphics[width=\textwidth]{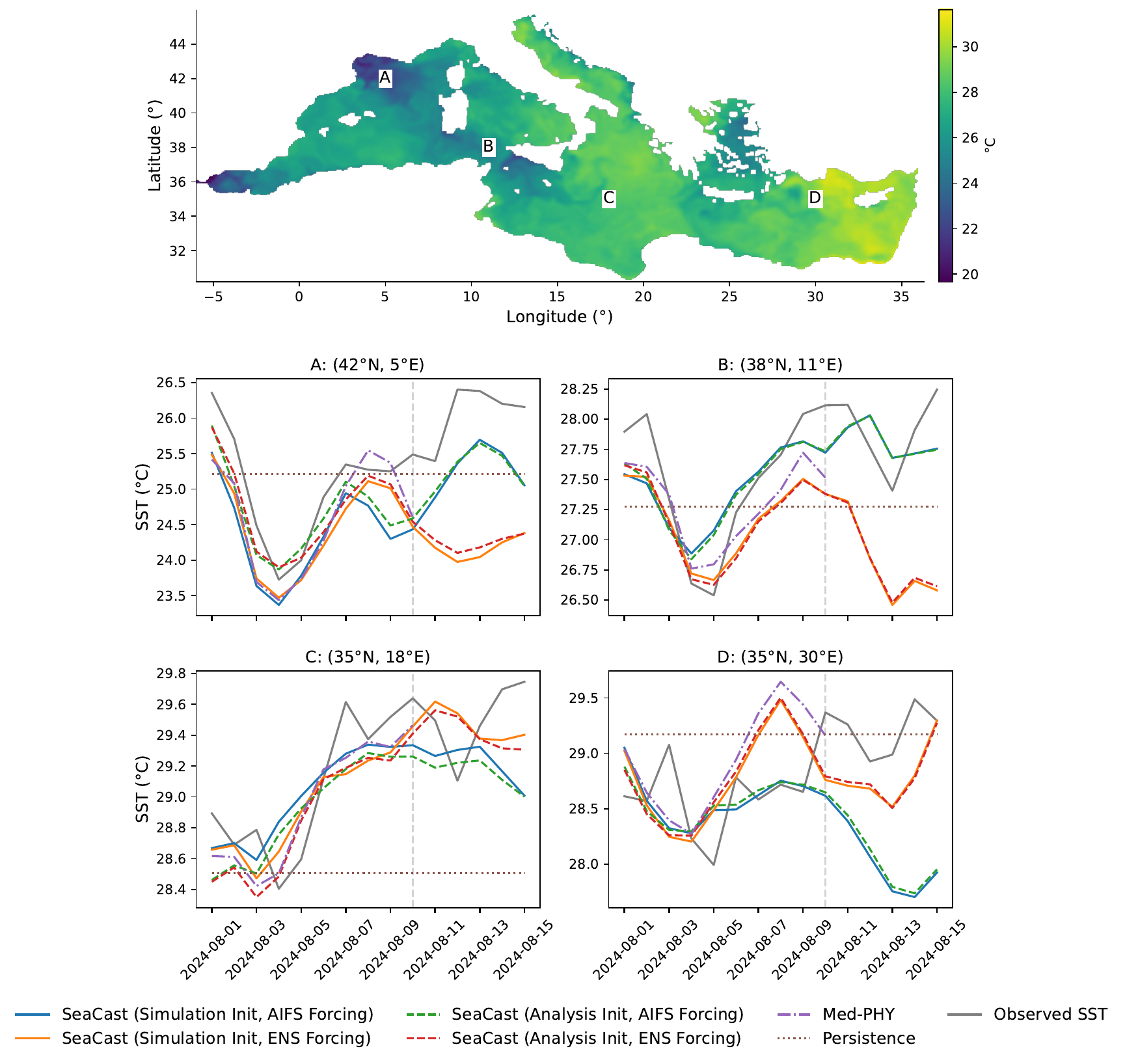}
    \caption{The top map displays satellite SST data from August 1st, 2024, with four labeled locations. The plots below show the evolution of the \texttt{thetao\_1} field at each location during a single forecast period, comparing the different models with satellite observation trends.}
    \label{fig:sst}
\end{figure}

For the Mediterranean Sea, the CNR MED SST product~\citeapp{cmems2024sst} provides daily gap-free sea surface temperature fields at a resolution of 1/16°. These datasets are produced in near-real time using nighttime infrared imagery from various satellite sensors. The SST data is bilinearly interpolated from its native resolution to the 1/24° resolution of the model outputs. Figure~\ref{fig:sst} shows the SST evolution over one forecast period starting on August 1st, 2024, at four locations, comparing the models' predictions of \texttt{thetao\_1} (the uppermost layer of temperature) with observed SST. The models capture SST trends quite well, particularly during the early days of the forecasts.

\begin{figure}[h!]
    \centering
    \includegraphics[width=.86\textwidth]{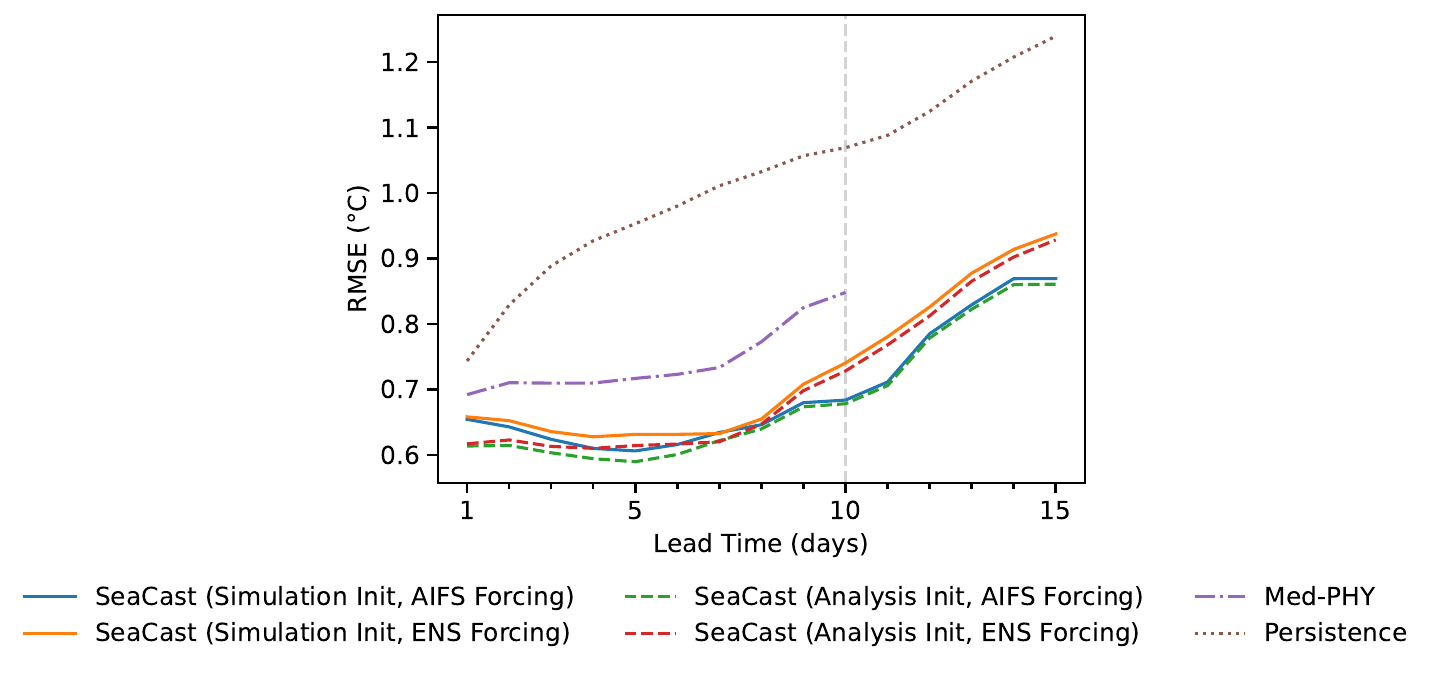}
    \caption{RMSE w.r.t. satellite SST for \texttt{thetao\_1} forecasts at different lead times.}
    \label{fig:rmse_sst}
\end{figure}

Arguably, one of the most interesting findings is shown in Figure~\ref{fig:rmse_sst}, which depicts the forecast RMSE for the entire SST test sample. Here, SeaCast outperforms Med-PHY by a margin. In the training data, both reanalysis and analysis datasets incorporate SST satellite information through a surface heat flux correction, which may help explain this phenomenon. We can further observe that the initialization plays a more significant role at the beginning, as models with analysis initialization have lower errors. In contrast, for longer lead times, atmospheric forcing becomes more influential, and models forced with AIFS data produce lower errors.

\begin{figure}[h]
    \centering
    \includegraphics[width=\textwidth]{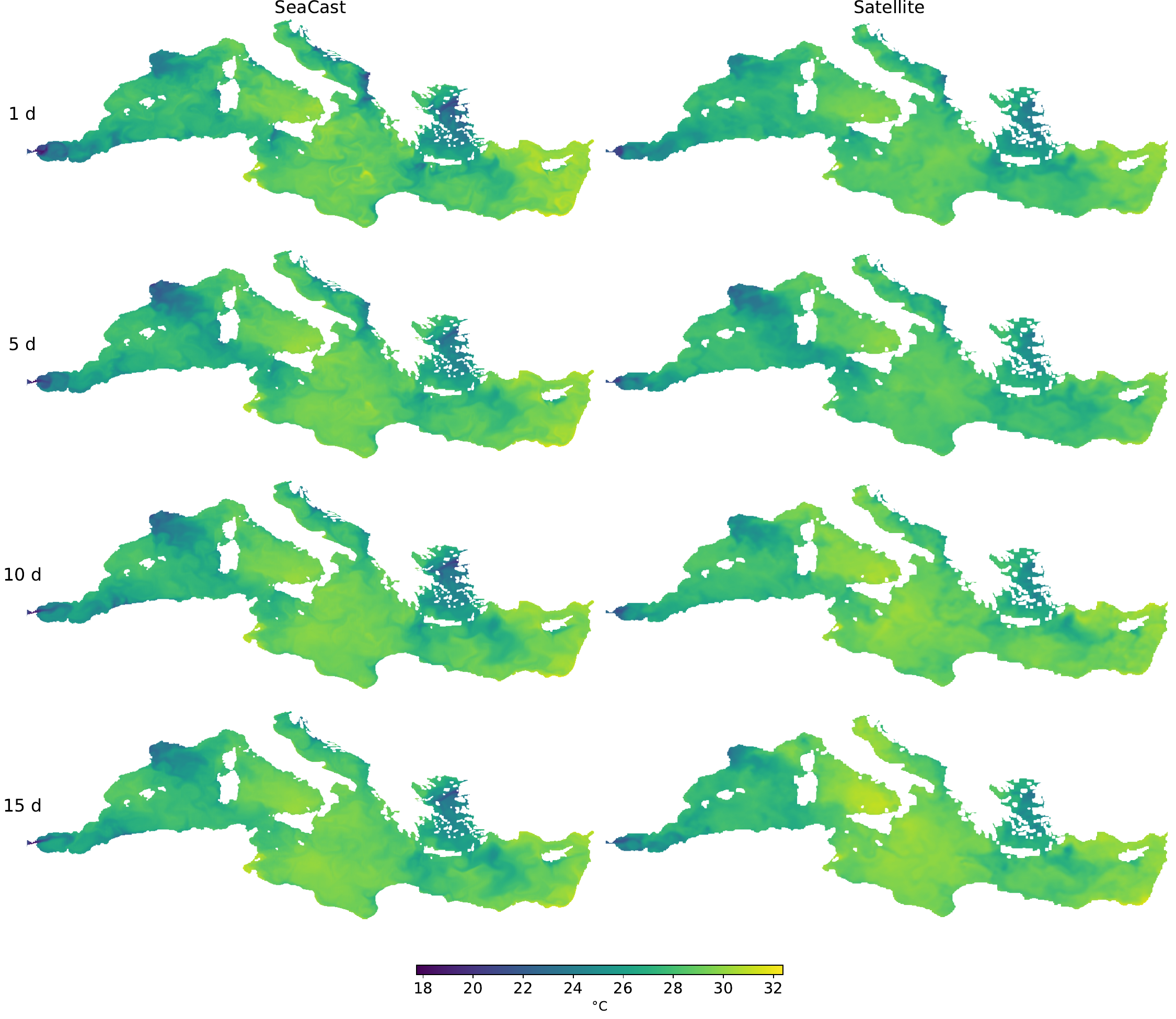}
    \caption{SeaCast-AIFS forecast for \texttt{thetao\_1} initialized on August 1st, 2024, vs. satellite SST.}
    \label{fig:rollout_sst}
\end{figure}

\begin{figure}[h]
    \centering
    \includegraphics[width=\textwidth]{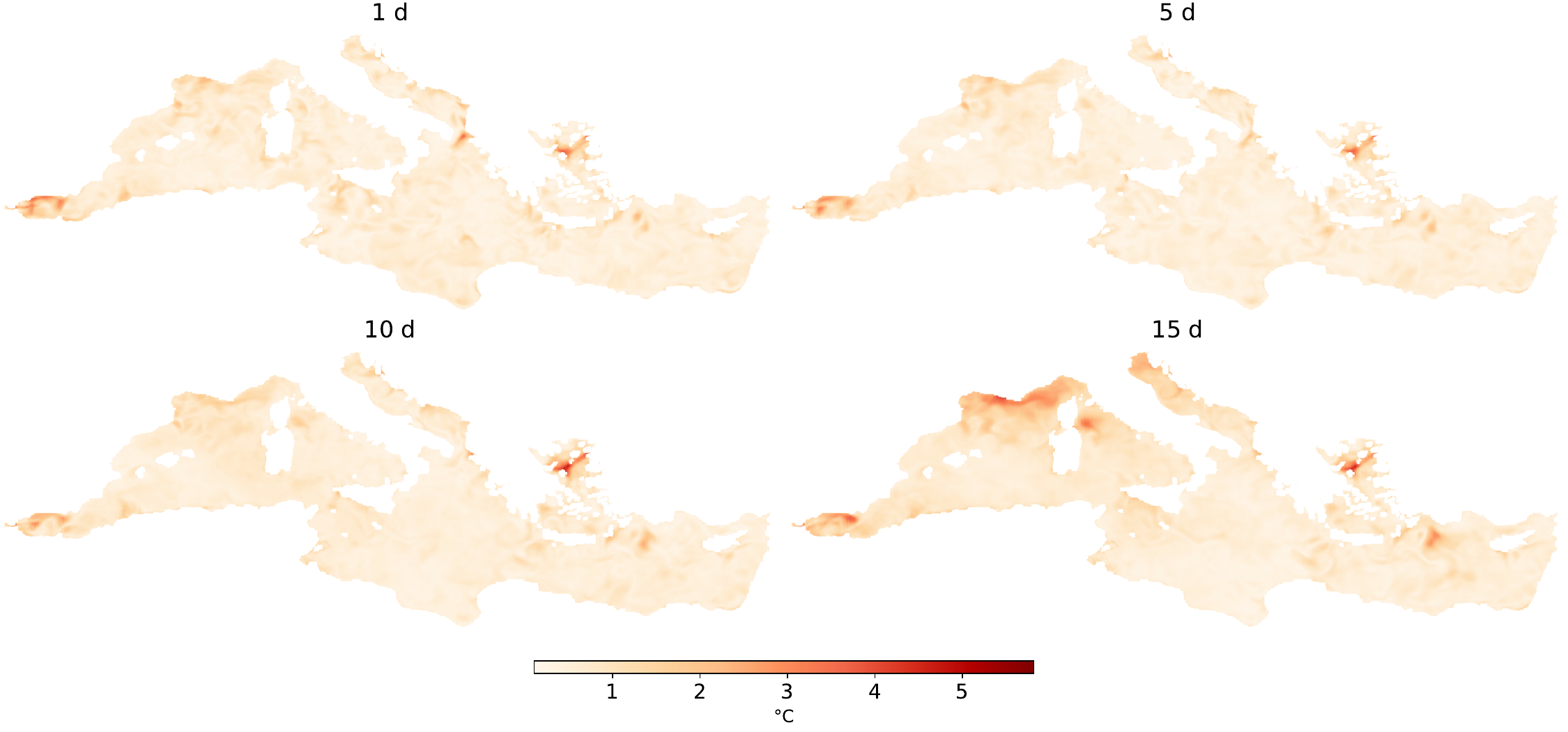}
    \caption{Spatial RMSE of SeaCast-AIFS \texttt{thetao\_1} forecasts vs. satellite SST at different leads.}
    \label{fig:error_sst}
\end{figure}

\begin{figure}[h]
    \centering
    \includegraphics[width=.82\textwidth]{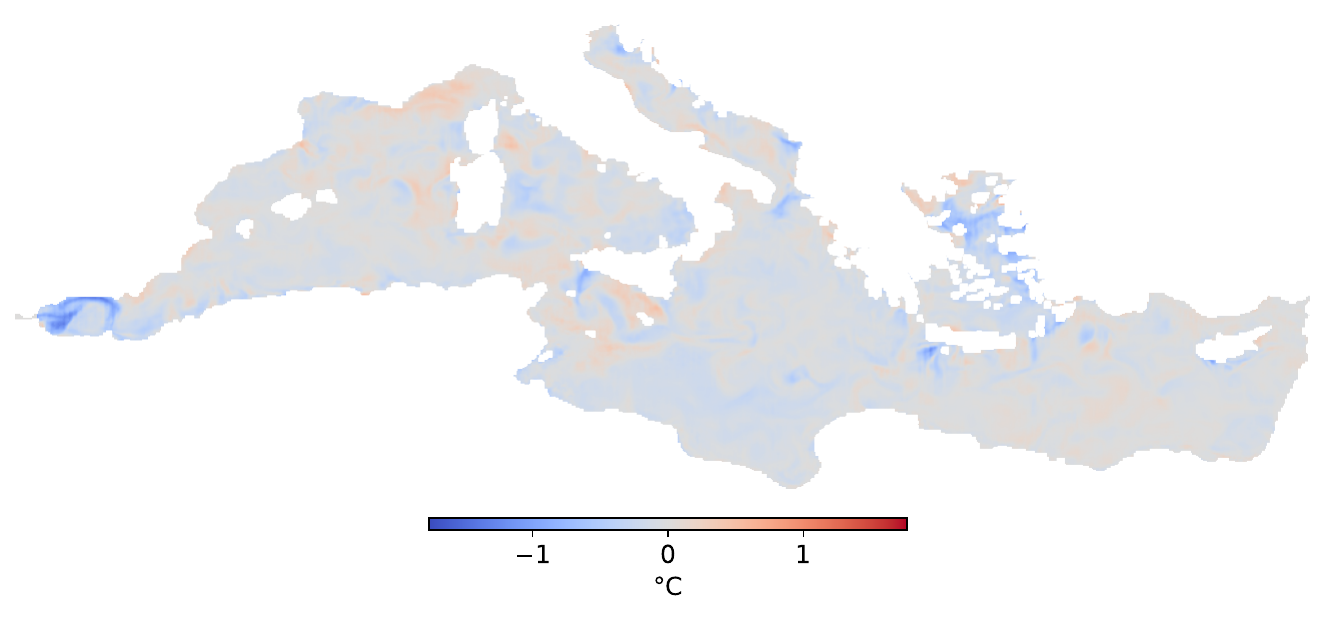}
    \caption{Spatial RMSE difference between SeaCast-ENS and Med-PHY \texttt{thetao\_1} forecasts compared to satellite SST. Blue indicates higher skill for SeaCast-ENS and red for Med-PHY.}
    \label{fig:bias_sst_seacast_ens_vs_med_phy}
\end{figure}

\begin{figure}[h]
    \centering
    \includegraphics[width=.82\textwidth]{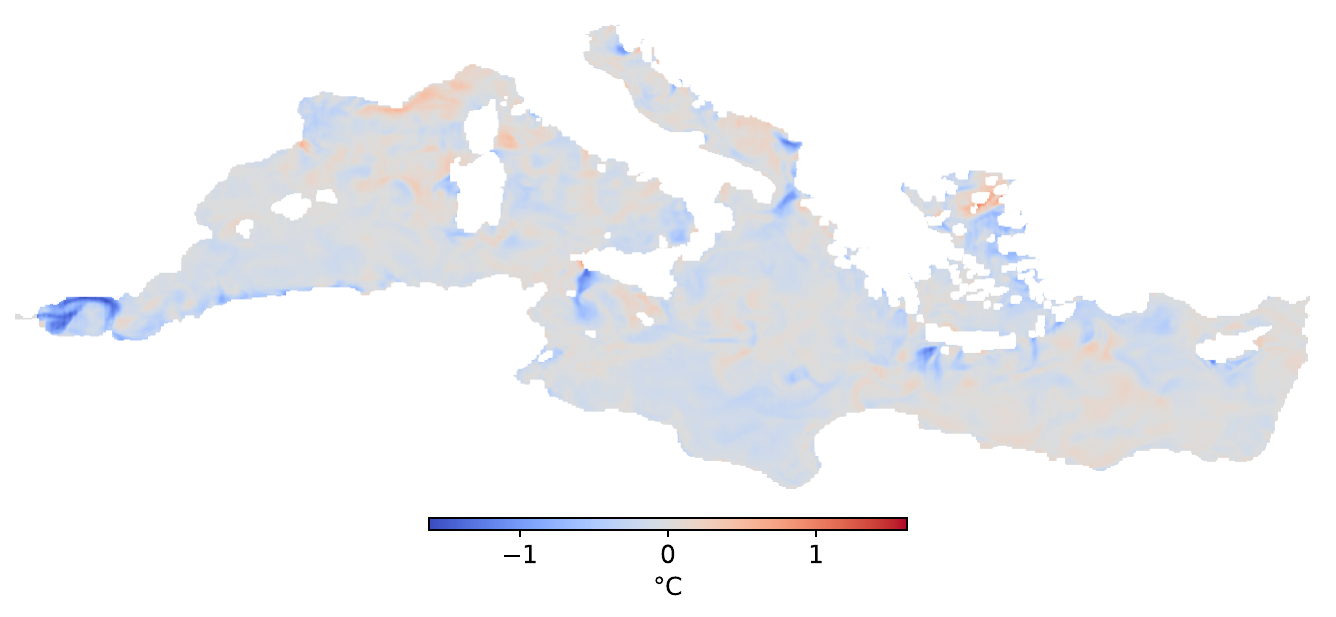}
    \caption{Spatial RMSE difference between SeaCast-AIFS and Med-PHY \texttt{thetao\_1} forecasts compared to satellite SST. Blue indicates higher skill for SeaCast-AIFS and red for Med-PHY.}
    \label{fig:bias_sst_seacast_aifs_vs_phy}
\end{figure}

\begin{figure}[h]
    \centering
    \includegraphics[width=.82\textwidth]{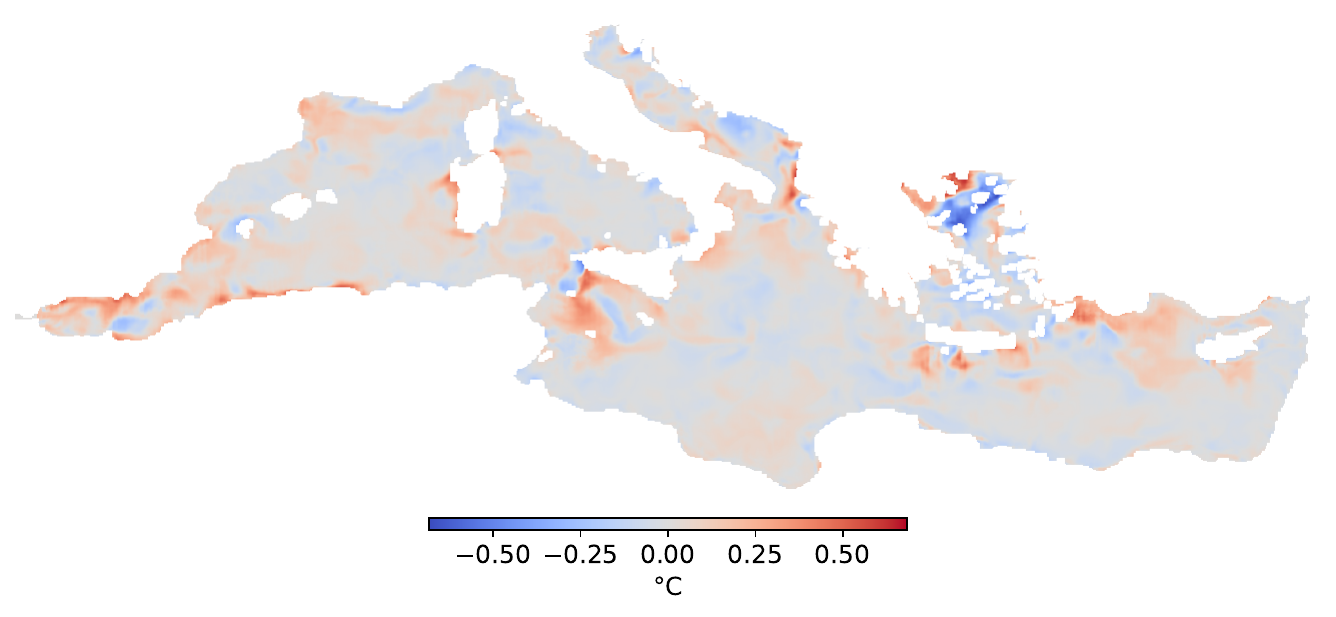}
    \caption{Spatial RMSE difference between SeaCast-ENS and SeaCast-AIFS \texttt{thetao\_1} forecasts compared to satellite SST. Blue indicates higher skill for SeaCast-ENS and red for SeaCast-AIFS.}
    \label{fig:bias_sst_seacast_ens_vs_aifs}
\end{figure}

Figure~\ref{fig:rollout_sst} presents the SeaCast-AIFS forecast of \texttt{thetao\_1} at four lead times compared to satellite observations, while Figure~\ref{fig:error_sst} shows the corresponding spatial errors averaged over all samples. Figures~\ref{fig:bias_sst_seacast_ens_vs_med_phy} to \ref{fig:bias_sst_seacast_ens_vs_aifs} depict the RMSE differences relative to satellite SST for the models, averaged over all samples up to a 10-day lead.

\subsection{Comparison to In Situ Measurements}

The CMEMS In Situ Thematic Assembly Centre (In Situ TAC) provides in situ observations from national and international observing systems~\citeapp{cmems2024insitu}. These data are essential for monitoring, modeling, and validation in forecasting systems, offering near real-time and reprocessed data products for the global ocean and regional seas.

\begin{figure}[h]
    \centering
    \includegraphics[width=.9\textwidth]{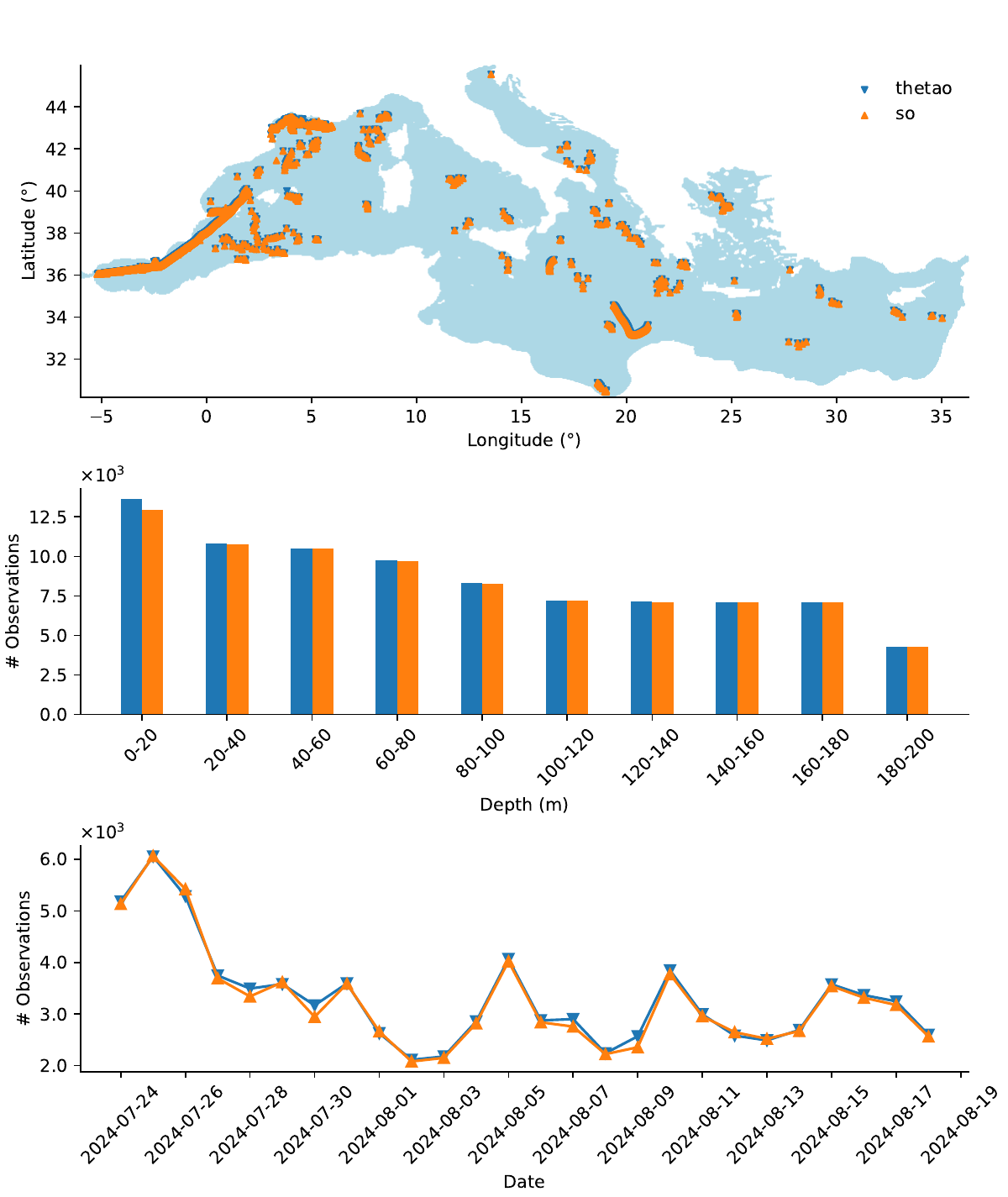}
    \caption{In situ observation statistics.}
    \label{fig:observations}
\end{figure}

\begin{figure}[h]
    \centering
    \includegraphics[width=.86\textwidth]{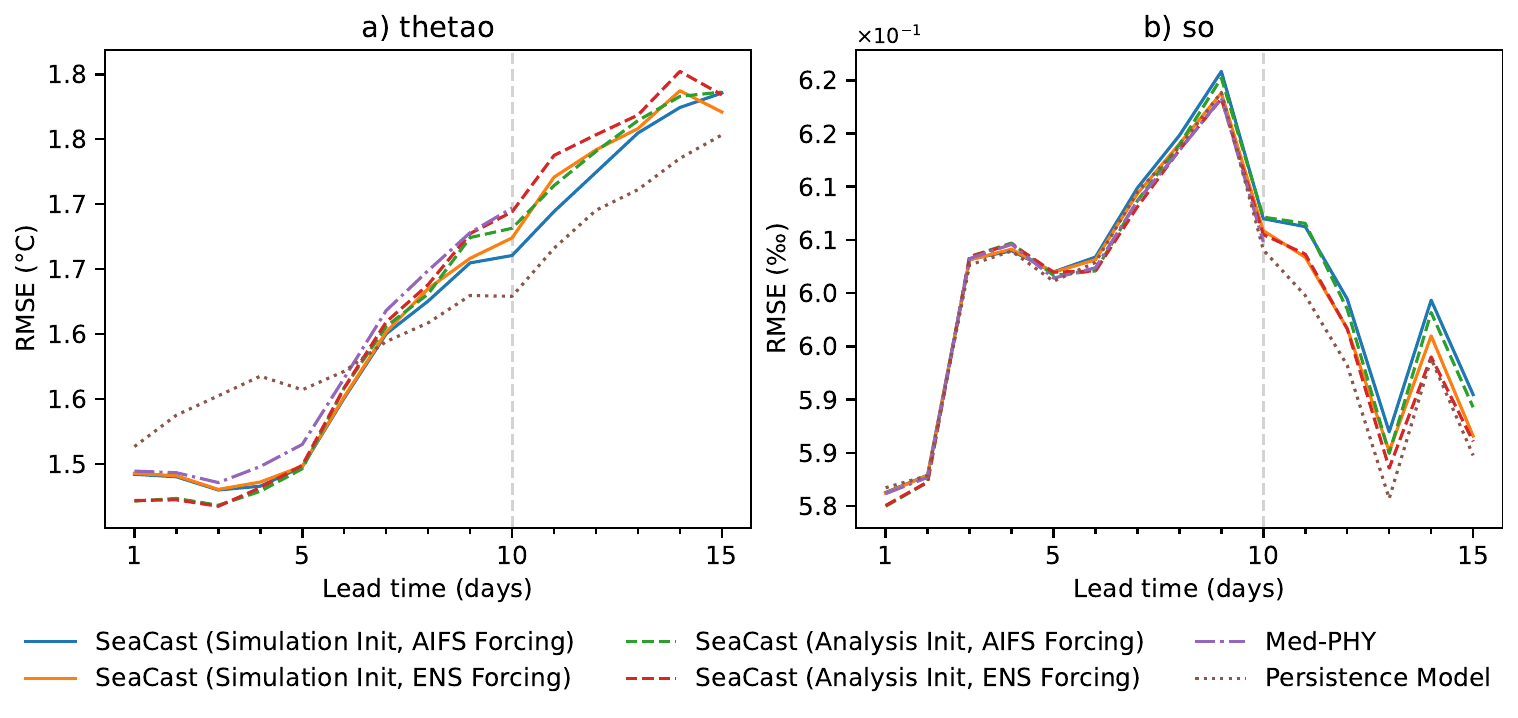}
    \caption{RMSE w.r.t. in situ observations for \texttt{thetao} and \texttt{so} forecasts at different lead times.}
    \label{fig:obs_rmse}
\end{figure}

Figure~\ref{fig:observations} shows the statistics of the downloaded in situ observations. To evaluate the performance of our models, we compared their outputs with in situ measurements of temperature (\texttt{thetao}) and salinity (\texttt{so}) by trilinearly interpolating the forecasted fields to the observation points, as depicted in Figure~\ref{fig:obs_rmse}. In this validation all models perform similarly. Measurements of zonal- and meridional velocity (\texttt{uo} and \texttt{vo}) were too sparse to allow for an effective evaluation during the given period.

\subsection{Comparison to Analysis}

\begin{figure}[h]
    \centering
    \includegraphics[width=\textwidth]{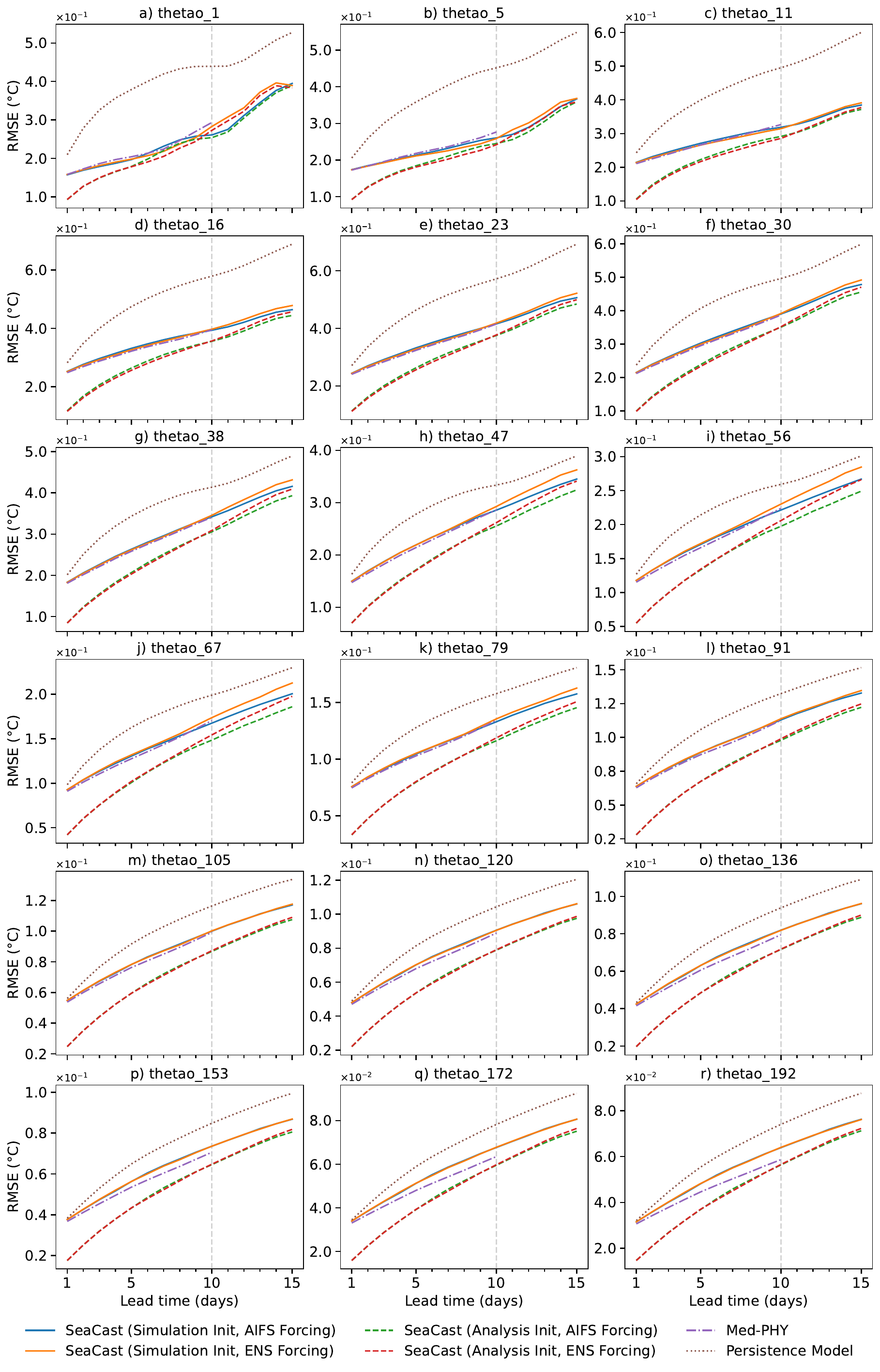}
    \caption{RMSE w.r.t. analysis fields for \texttt{thetao} forecasts at different lead times.}
    \label{fig:rmse_thetao}
\end{figure}

\begin{figure}[h]
    \centering
    \includegraphics[width=\textwidth]{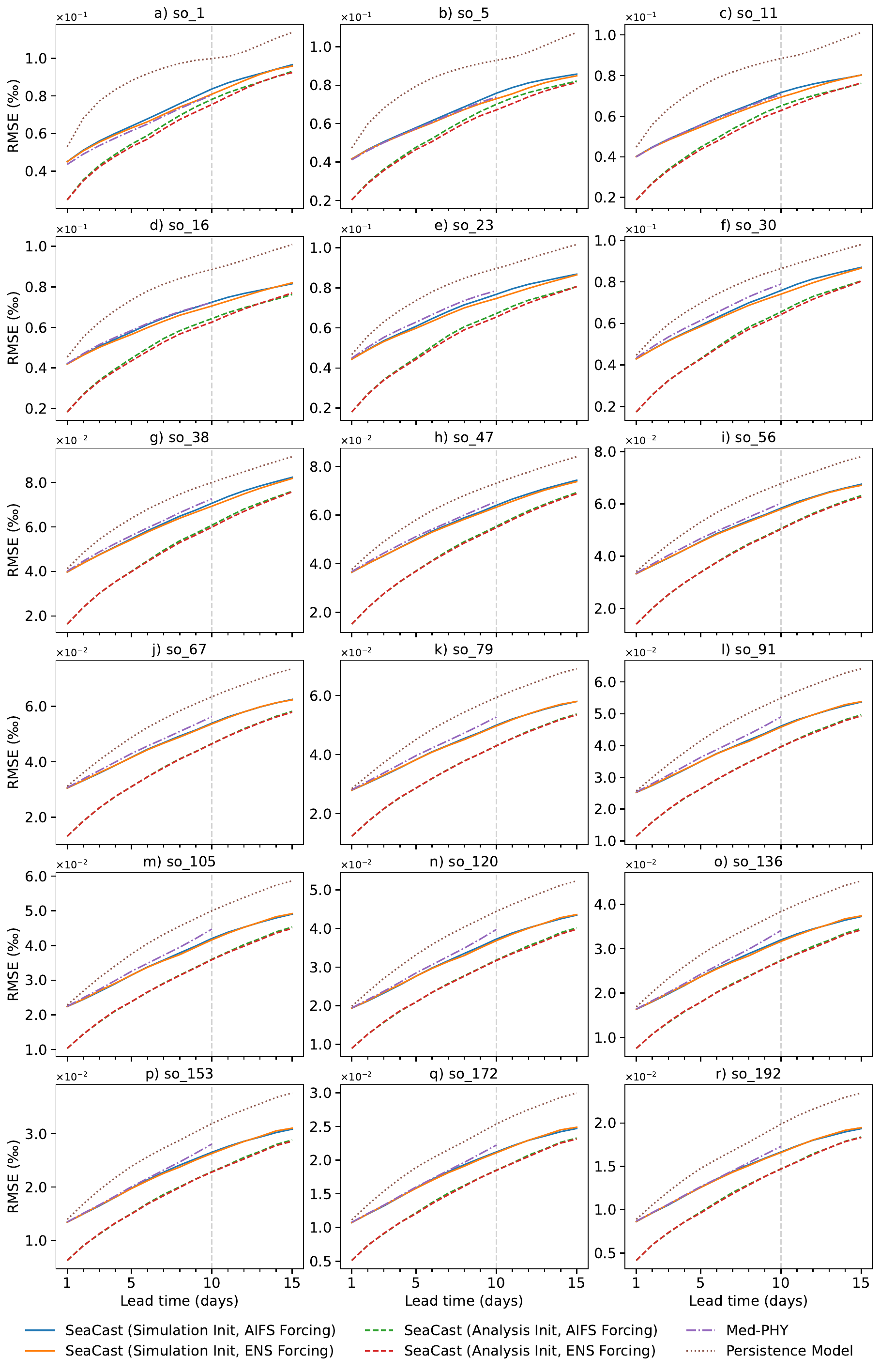}
    \caption{RMSE w.r.t. analysis fields for \texttt{so} forecasts at different lead times.}
    \label{fig:rmse_so}
\end{figure}

\begin{figure}[h]
    \centering
    \includegraphics[width=\textwidth]{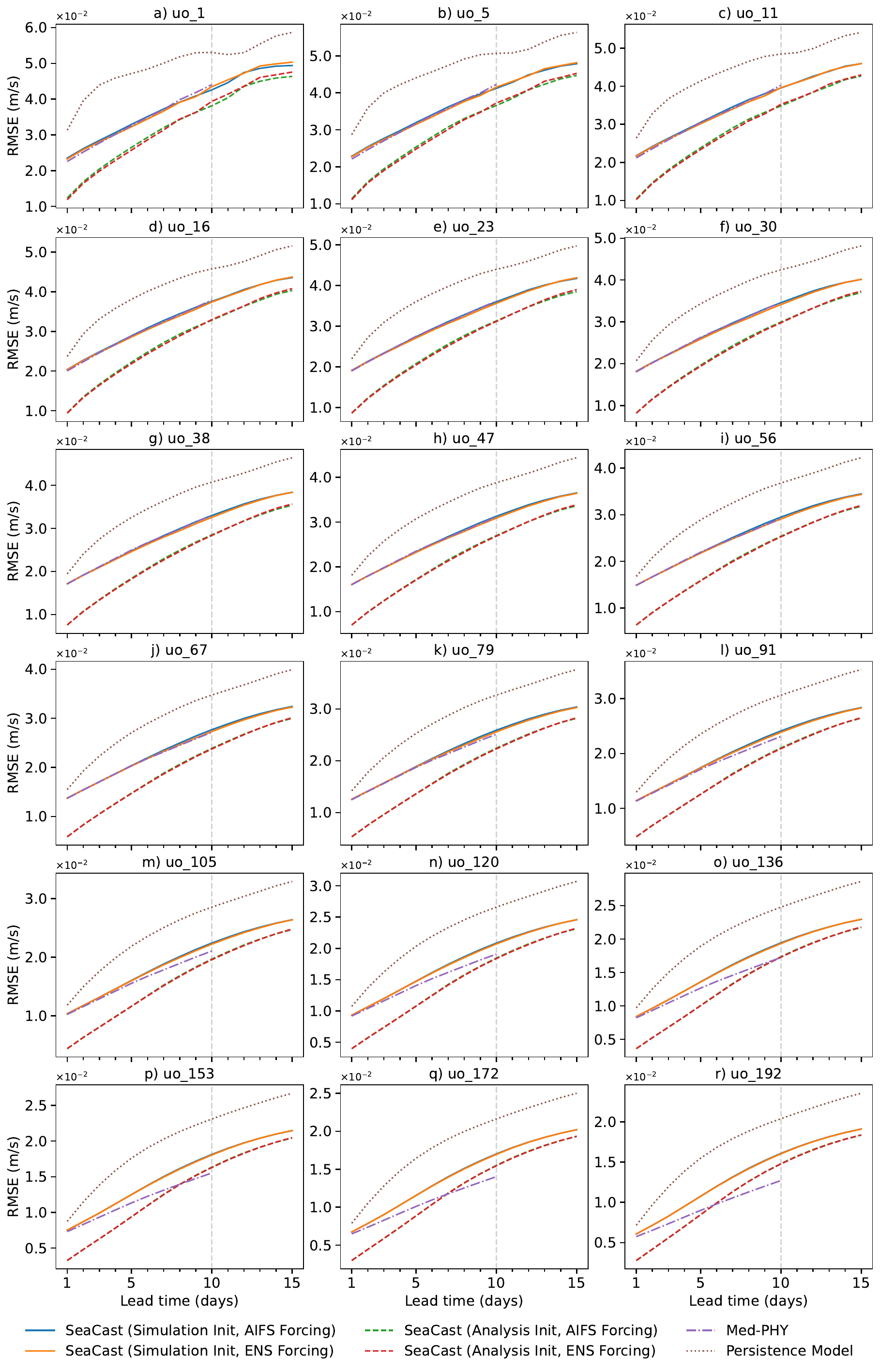}
    \caption{RMSE w.r.t. analysis fields for \texttt{uo} forecasts at different lead times.}
    \label{fig:rmse_uo}
\end{figure}

\begin{figure}[h]
    \centering
    \includegraphics[width=\textwidth]{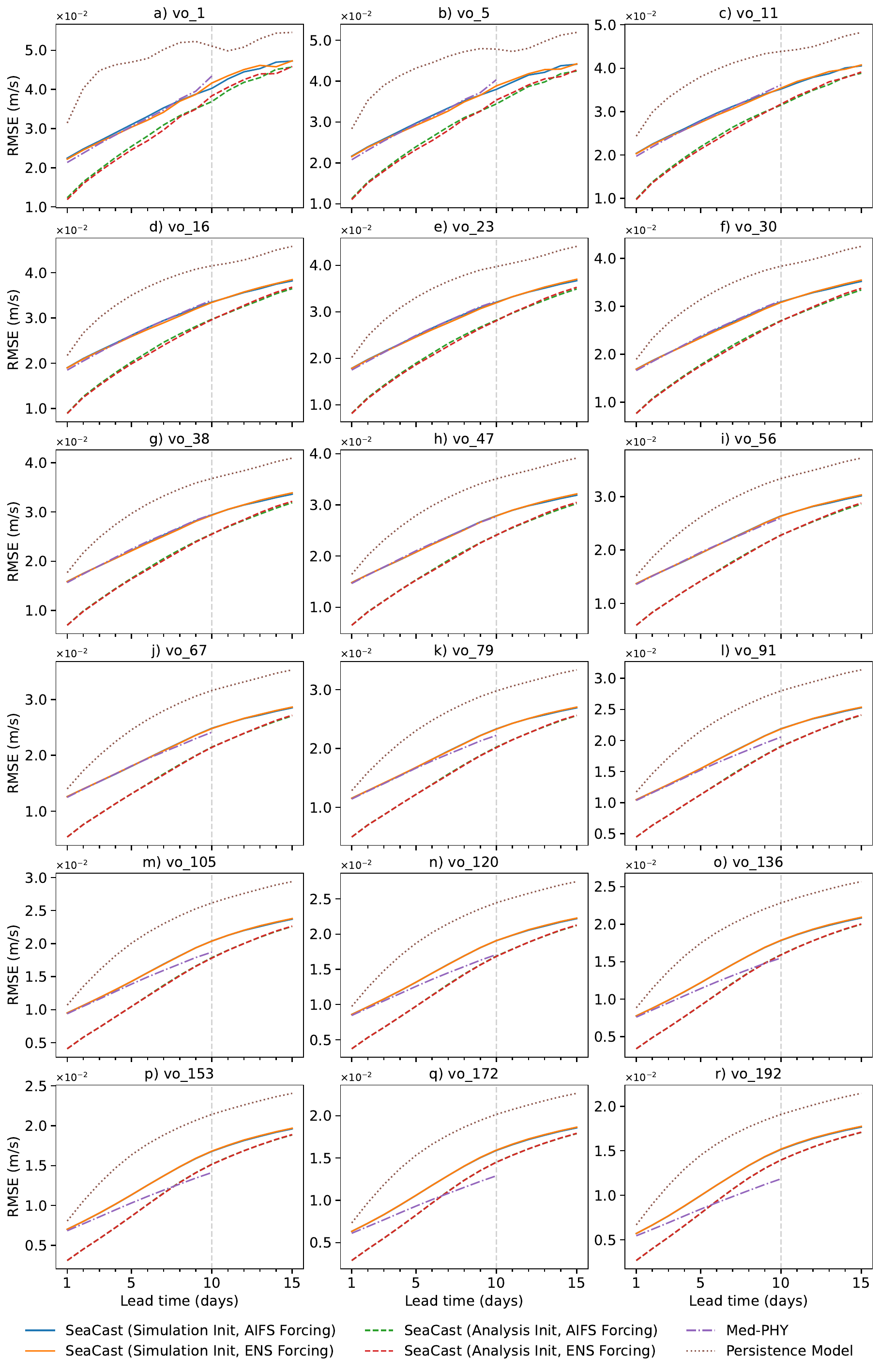}
    \caption{RMSE w.r.t. analysis fields for \texttt{vo} forecasts at different lead times.}
    \label{fig:rmse_vo}
\end{figure}

\begin{figure}[h]
    \centering
    \includegraphics[width=\textwidth]{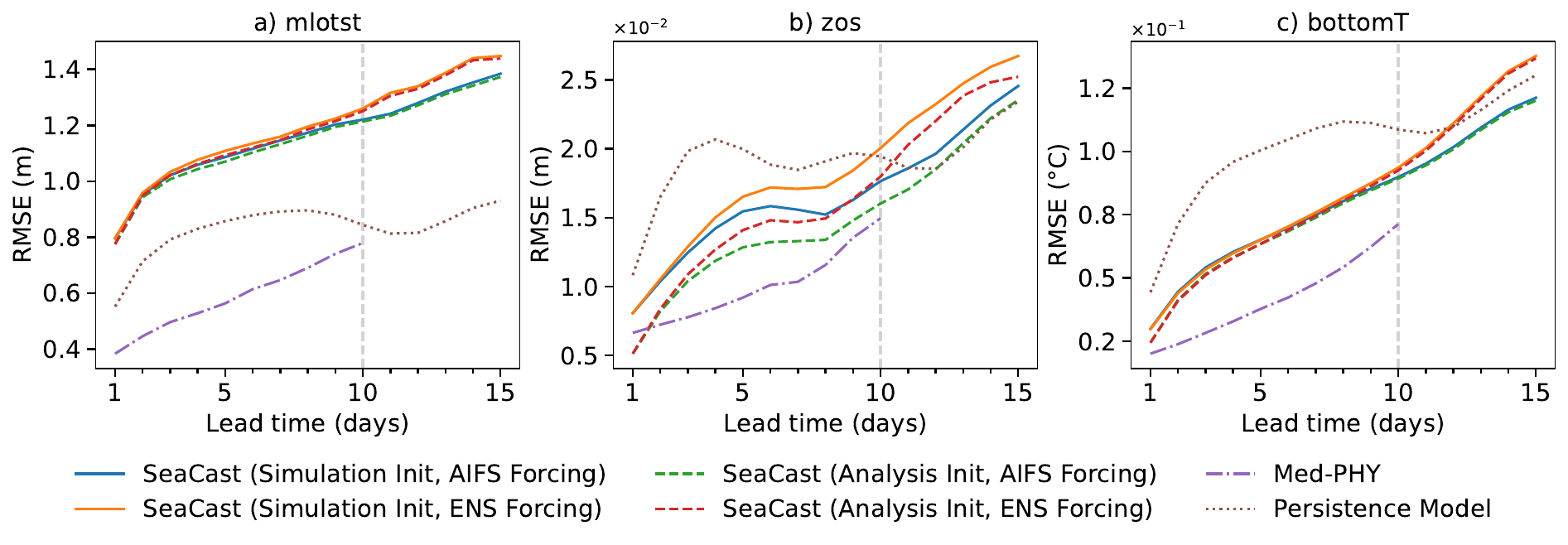}
    \caption{RMSE w.r.t. analysis fields for \texttt{mlotst}, \texttt{zos}, and \texttt{bottomT} at different lead times.}
    \label{fig:rmse_single_lev}
\end{figure}

The average RMSE across samples is calculated for all forecasted fields over all lead times, as shown in Figures~\ref{fig:rmse_thetao} to \ref{fig:rmse_single_lev}. The SeaCast models initialized from operational simulation states outperform the persistence model and perform on par with Med-PHY. It is worth noting that the ML models effectively have an open boundary below 200 m, as fields deeper than that are not included in the training data. This leads to slightly larger errors compared to Med-PHY with respect to analysis fields at depths of roughly 80 m and below. This effect is evident in temperature and velocity forecasts but does not appear to affect salinity forecasts.

\begin{figure}[h]
    \centering
    \includegraphics[width=.77\textwidth]{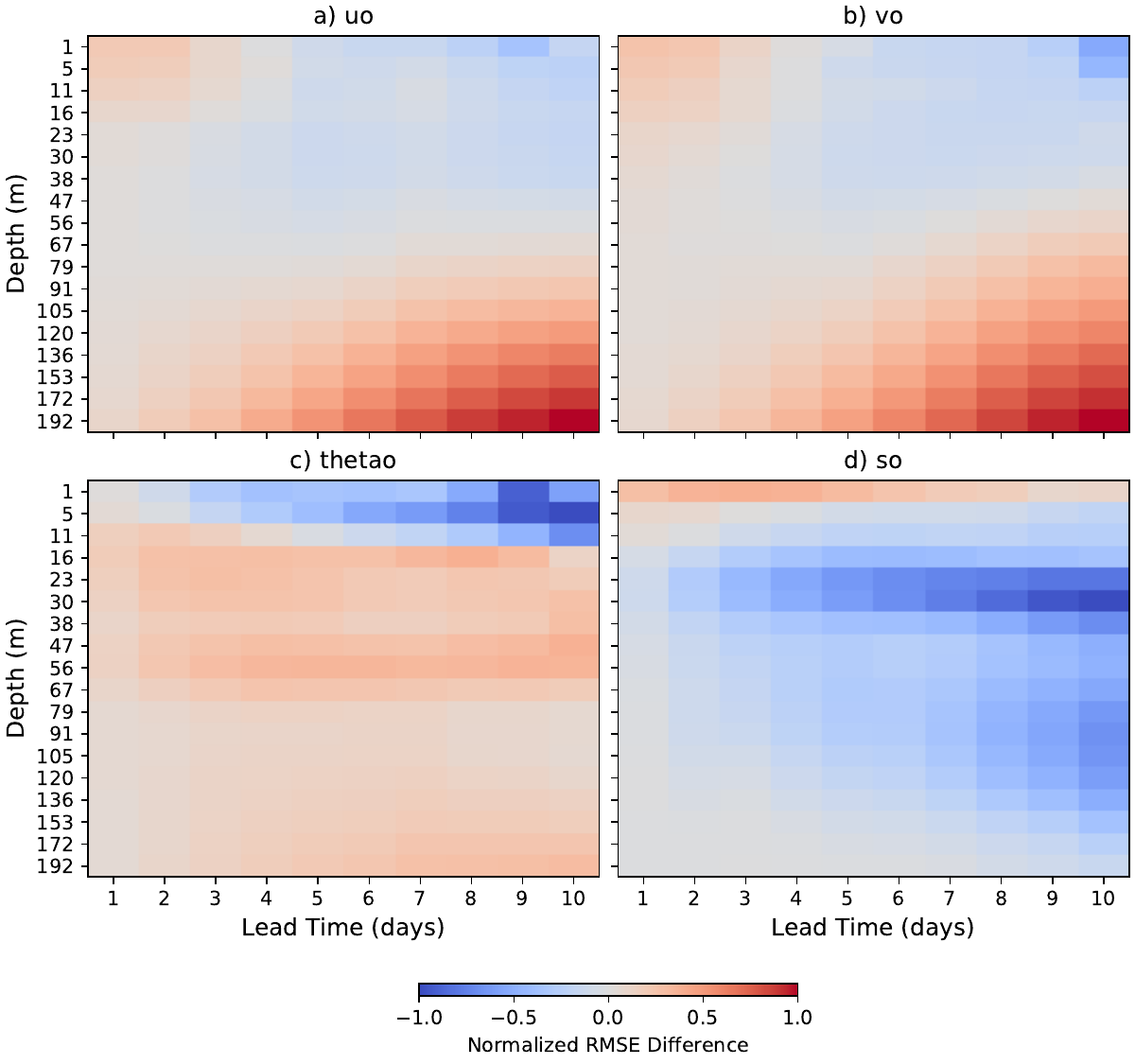}
    \caption{Normalized RMSE difference between SeaCast-ENS and Med-PHY compared to analysis. Blue indicates higher skill for SeaCast-ENS, and red for Med-PHY.}
    \label{fig:heatmap_seacast_ens_vs_med_phy}
\end{figure}

\begin{figure}[h]
    \centering
    \includegraphics[width=.77\textwidth]{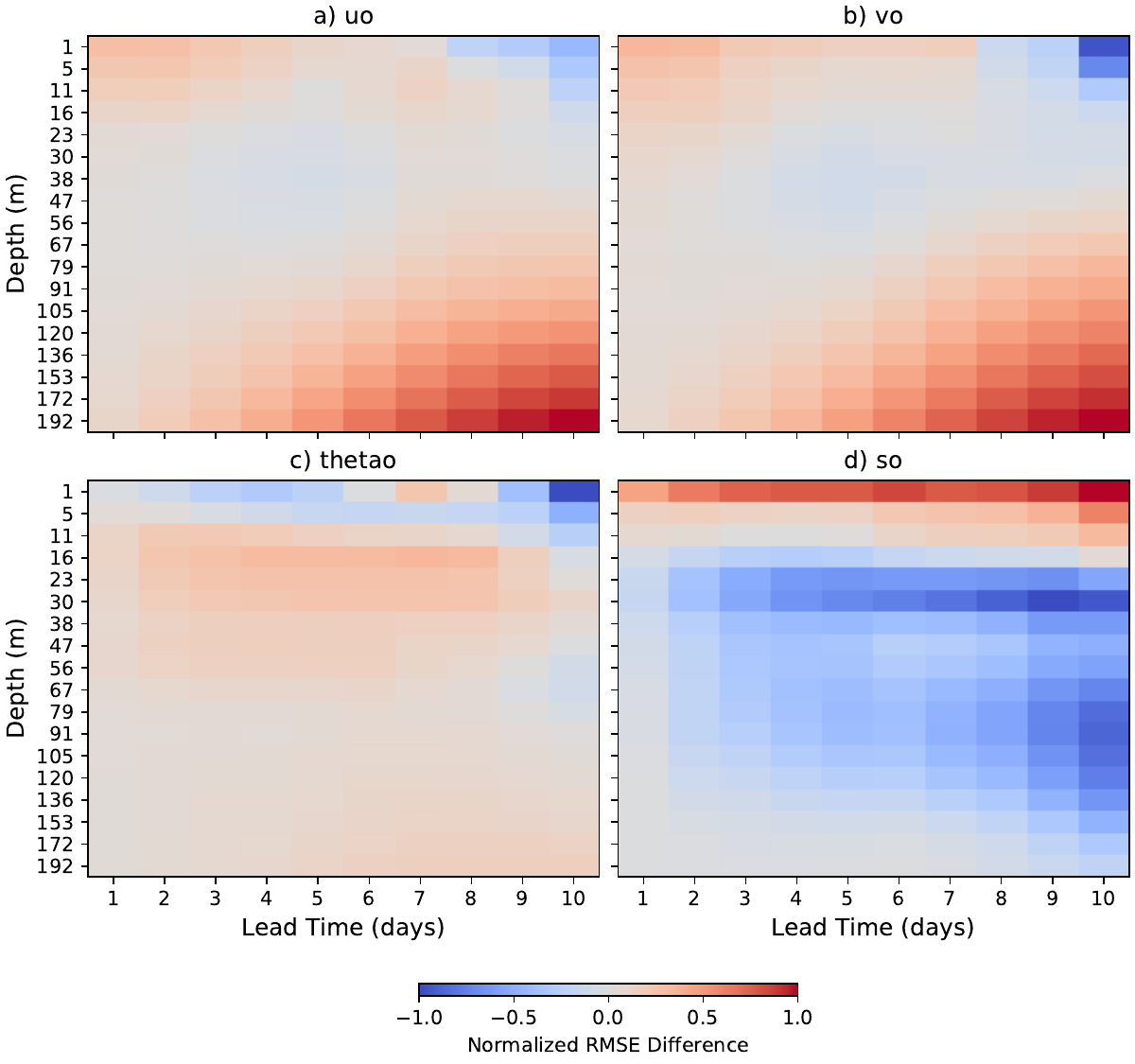}
    \caption{Normalized RMSE difference between SeaCast-AIFS and Med-PHY compared to analysis. Blue indicates higher skill for SeaCast-AIFS, and red for Med-PHY.}
    \label{fig:heatmap_seacast_aifs_vs_med_phy}
\end{figure}

\begin{figure}[h]
    \centering
    \includegraphics[width=.77\textwidth]{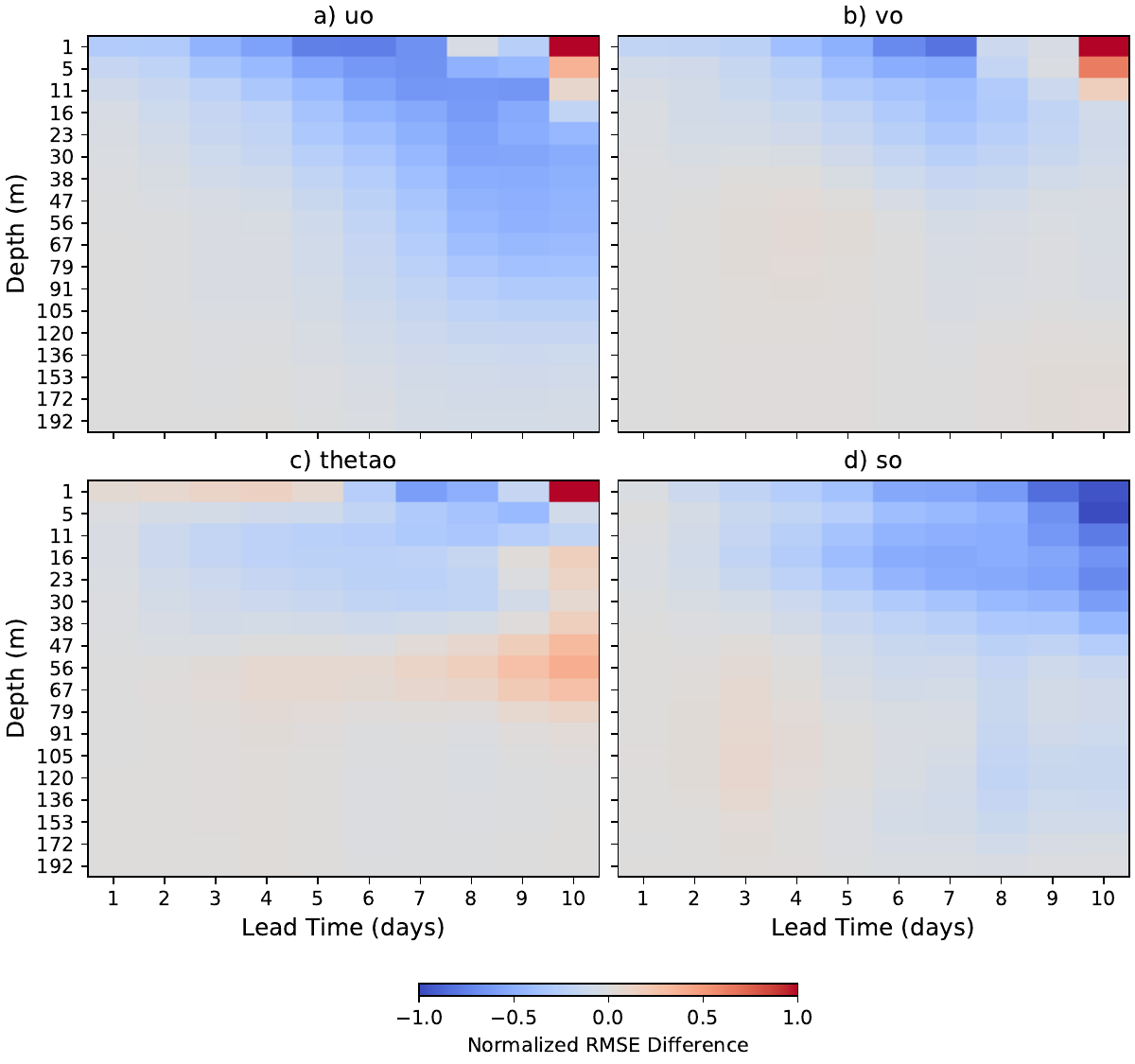}
    \caption{Normalized RMSE difference between SeaCast-ENS and SeaCast-AIFS compared to analysis. Blue indicates higher skill for SeaCast-ENS, and red for SeaCast-AIFS.}
    \label{fig:heatmap_seacast_ens_vs_aifs}
\end{figure}

Figures~\ref{fig:heatmap_seacast_ens_vs_med_phy} to \ref{fig:heatmap_seacast_ens_vs_aifs} show the RMSE differences between the models when compared to analysis fields. Generally, the ML models produce lower errors than Med-PHY for upper-layer temperature, velocity, and salinity overall. However, as observed in the previous RMSE line plots over time, these differences are small, even though they appear more pronounced in the heatmaps due to normalization.

\begin{figure}[h]
    \centering
    \includegraphics[width=\textwidth]{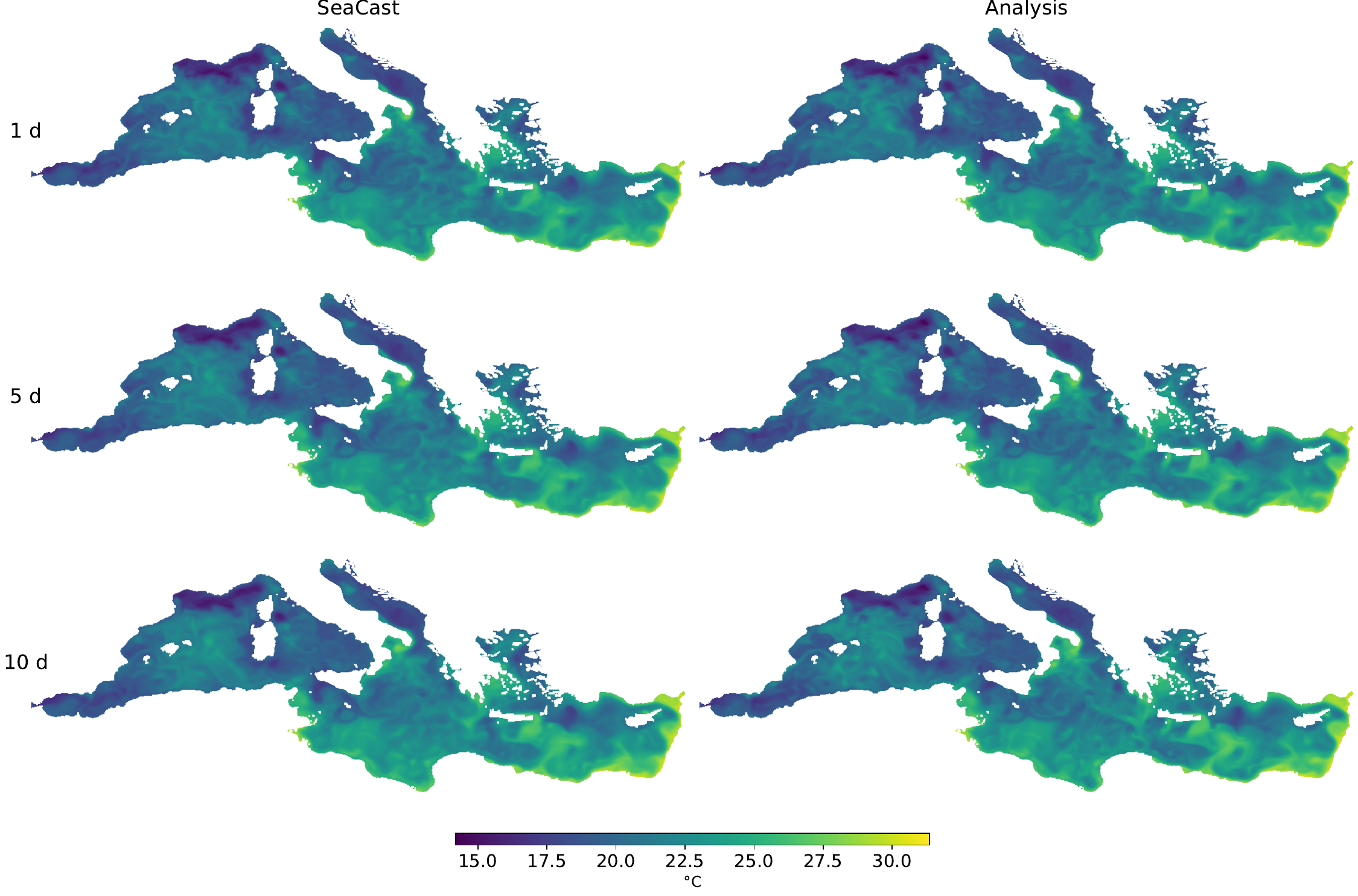}
    \caption{SeaCast-AIFS forecast for \texttt{thetao\_30} initialized on August 1st, 2024, vs. analysis.}
    \label{fig:rollout_thetao_30}
\end{figure}

\begin{figure}[h]
    \centering
    \includegraphics[width=\textwidth]{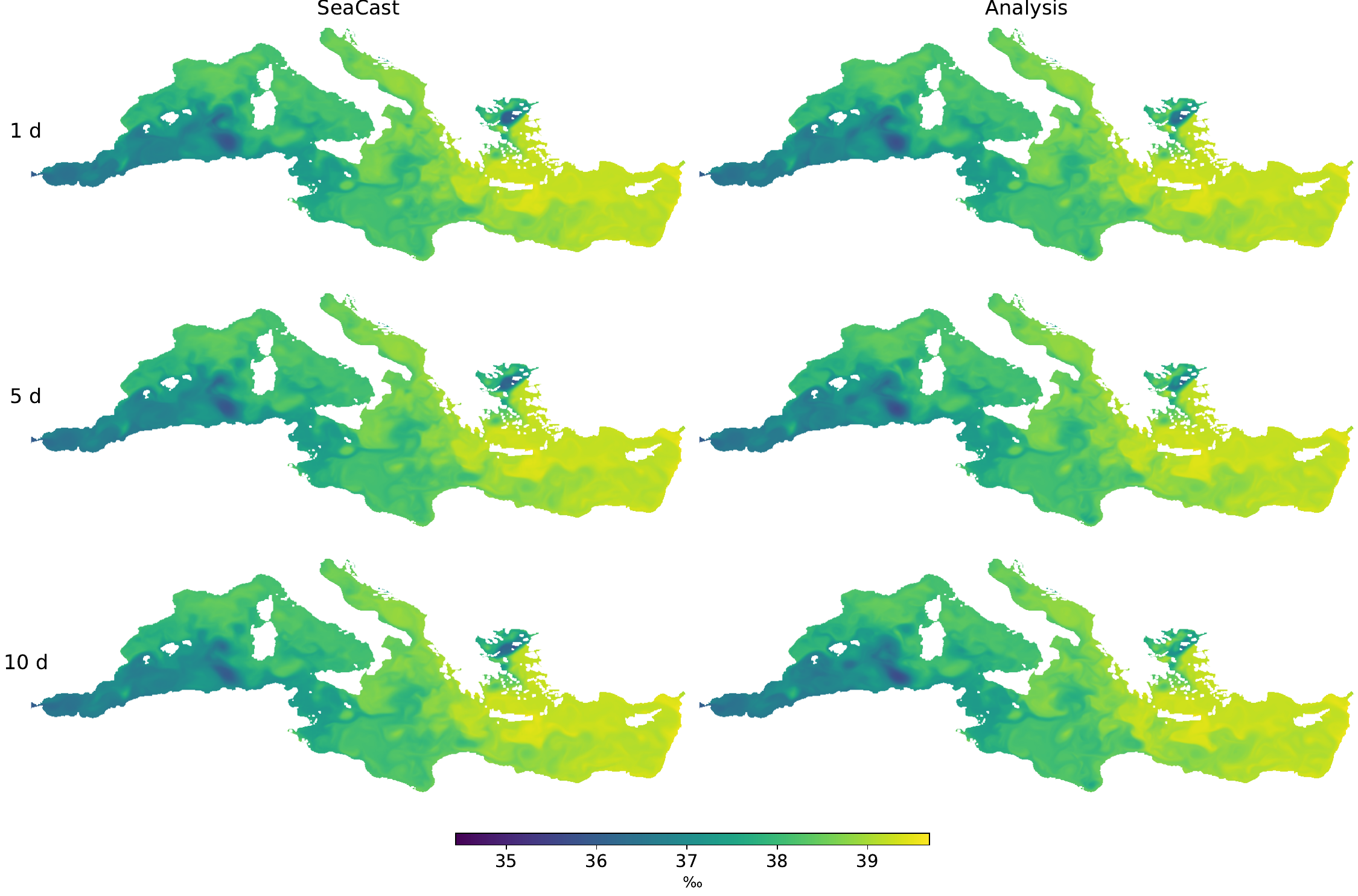}
    \caption{SeaCast-AIFS forecast for \texttt{so\_30} initialized on August 1st, 2024, vs. analysis.}
    \label{fig:rollout_so_30}
\end{figure}

\begin{figure}[h]
    \centering
    \includegraphics[width=\textwidth]{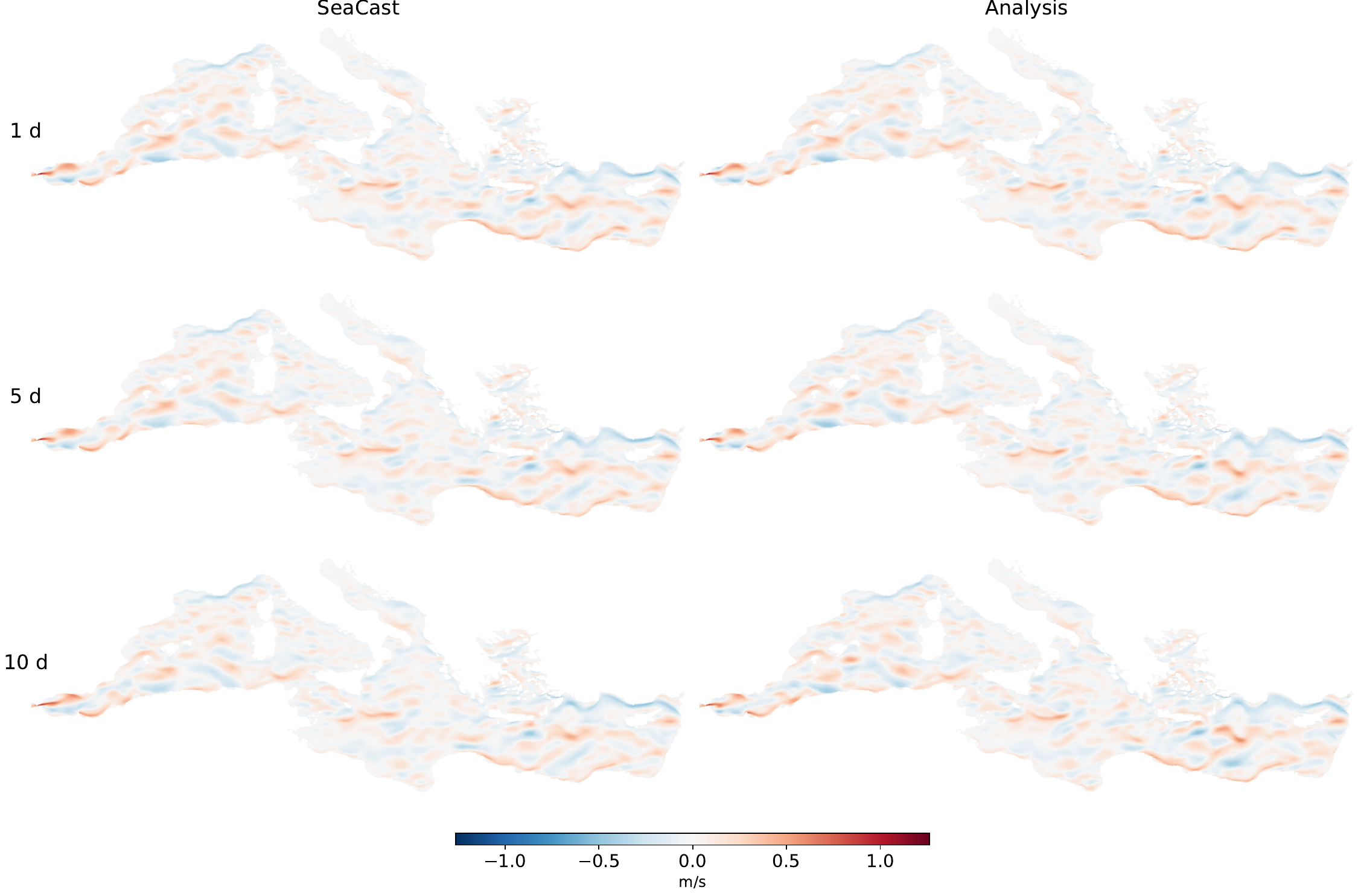}
    \caption{SeaCast-AIFS forecast for \texttt{uo\_30} initialized on August 1st, 2024, vs. analysis.}
    \label{fig:rollout_uo_30}
\end{figure}

\begin{figure}[h]
    \centering
    \includegraphics[width=\textwidth]{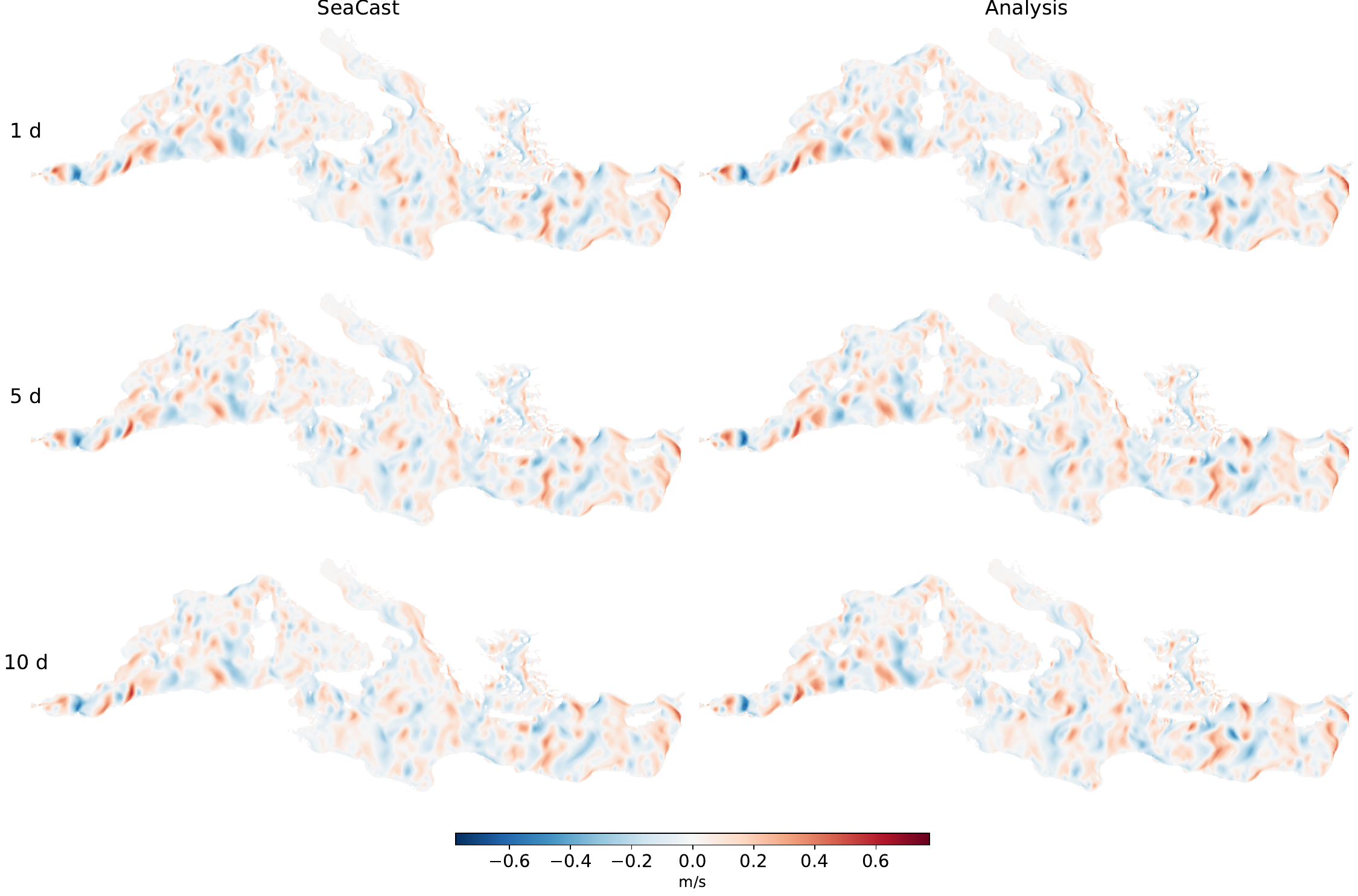}
    \caption{SeaCast-AIFS forecast for \texttt{vo\_30} initialized on August 1st, 2024, vs. analysis.}
    \label{fig:rollout_vo_30}
\end{figure}

\begin{figure}[h]
    \centering
    \includegraphics[width=\textwidth]{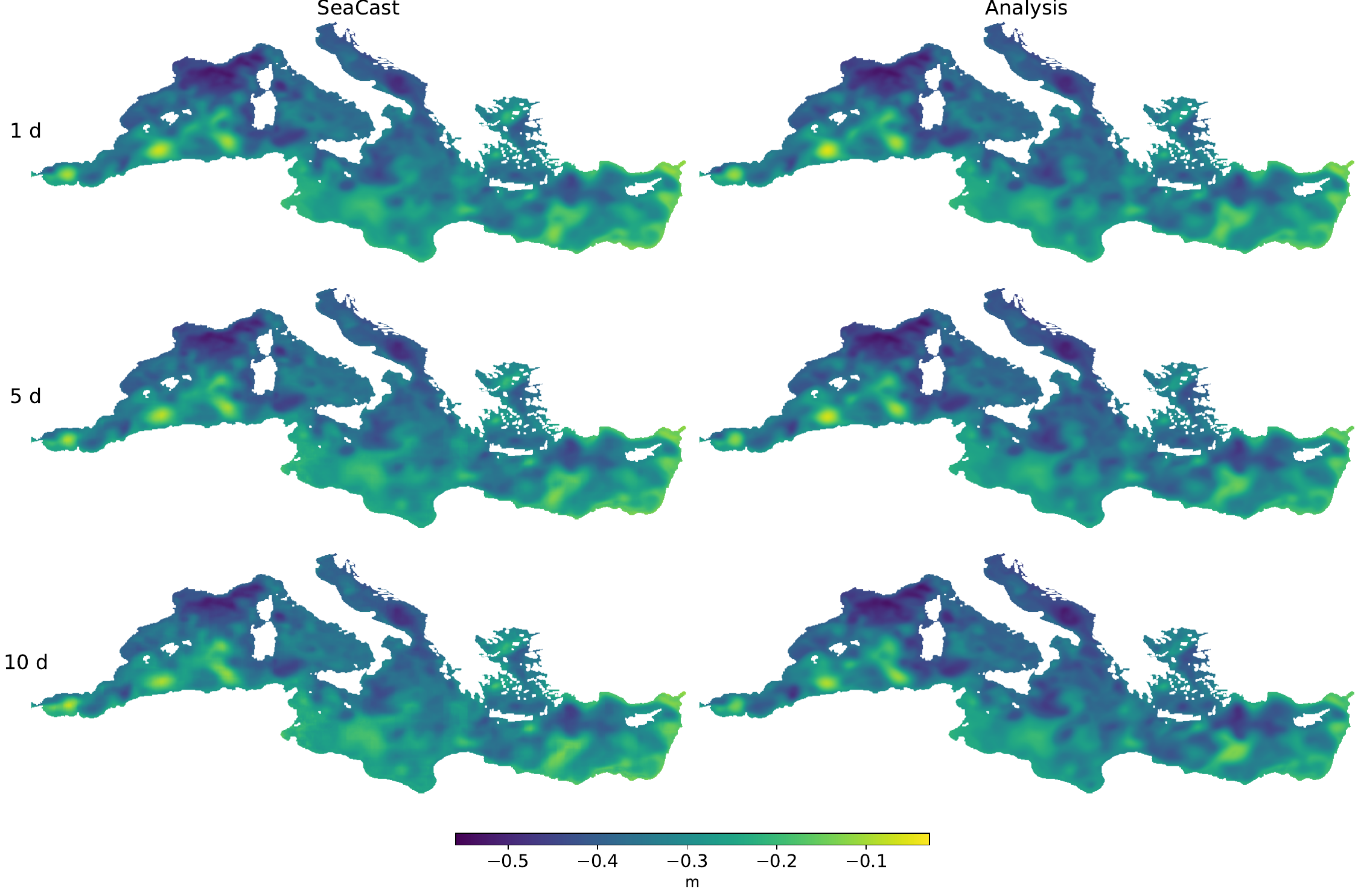}
    \caption{SeaCast-AIFS forecast for \texttt{zos} initialized on August 1st, 2024, vs. analysis.}
    \label{fig:rollout_zos}
\end{figure}

\begin{figure}[h]
    \centering
    \includegraphics[width=\textwidth]{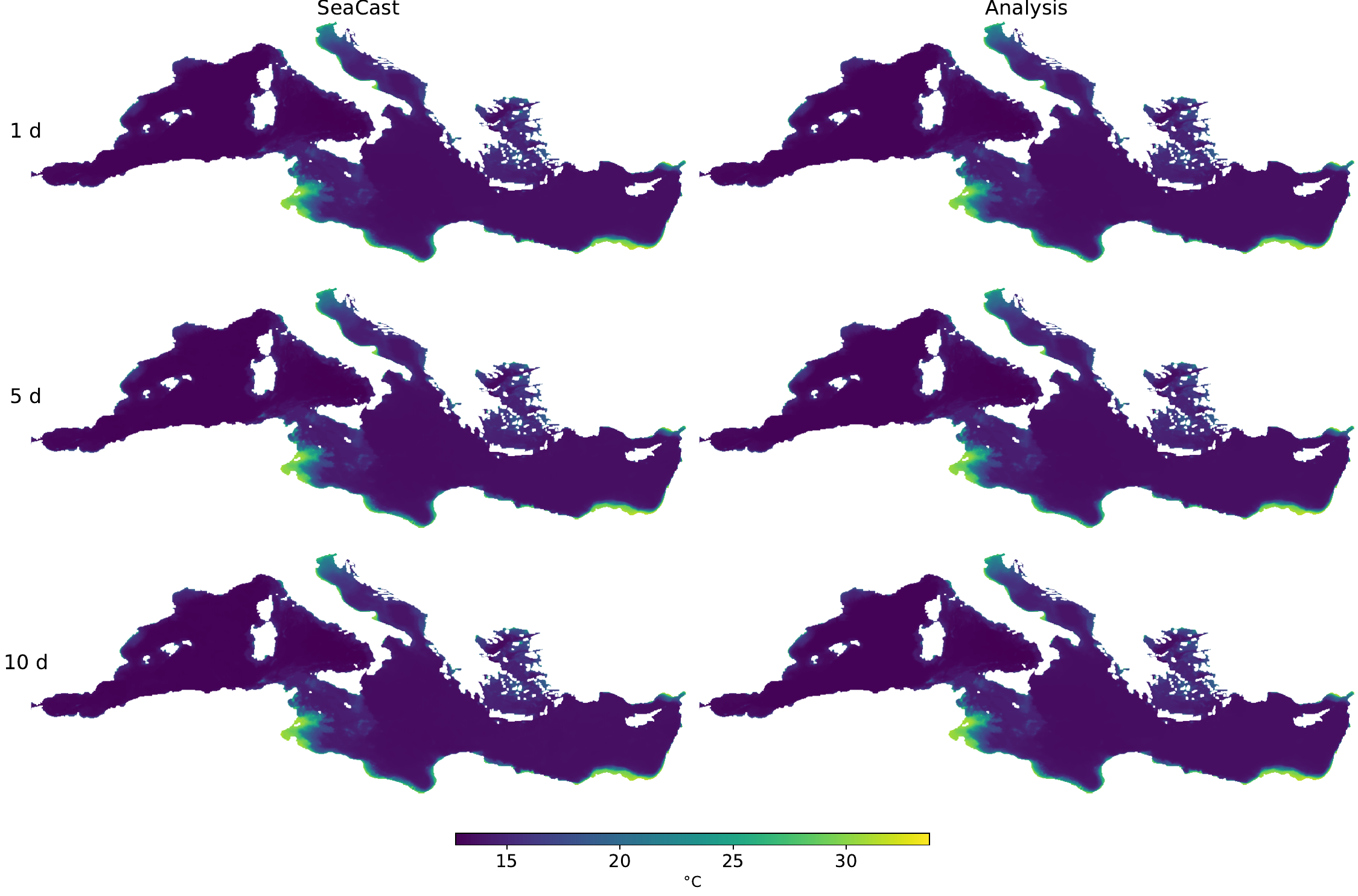}
    \caption{SeaCast-AIFS forecast for \texttt{bottomT} initialized on August 1st, 2024, vs. analysis.}
    \label{fig:rollout_bottomT}
\end{figure}

\begin{figure}[h]
    \centering
    \includegraphics[width=\textwidth]{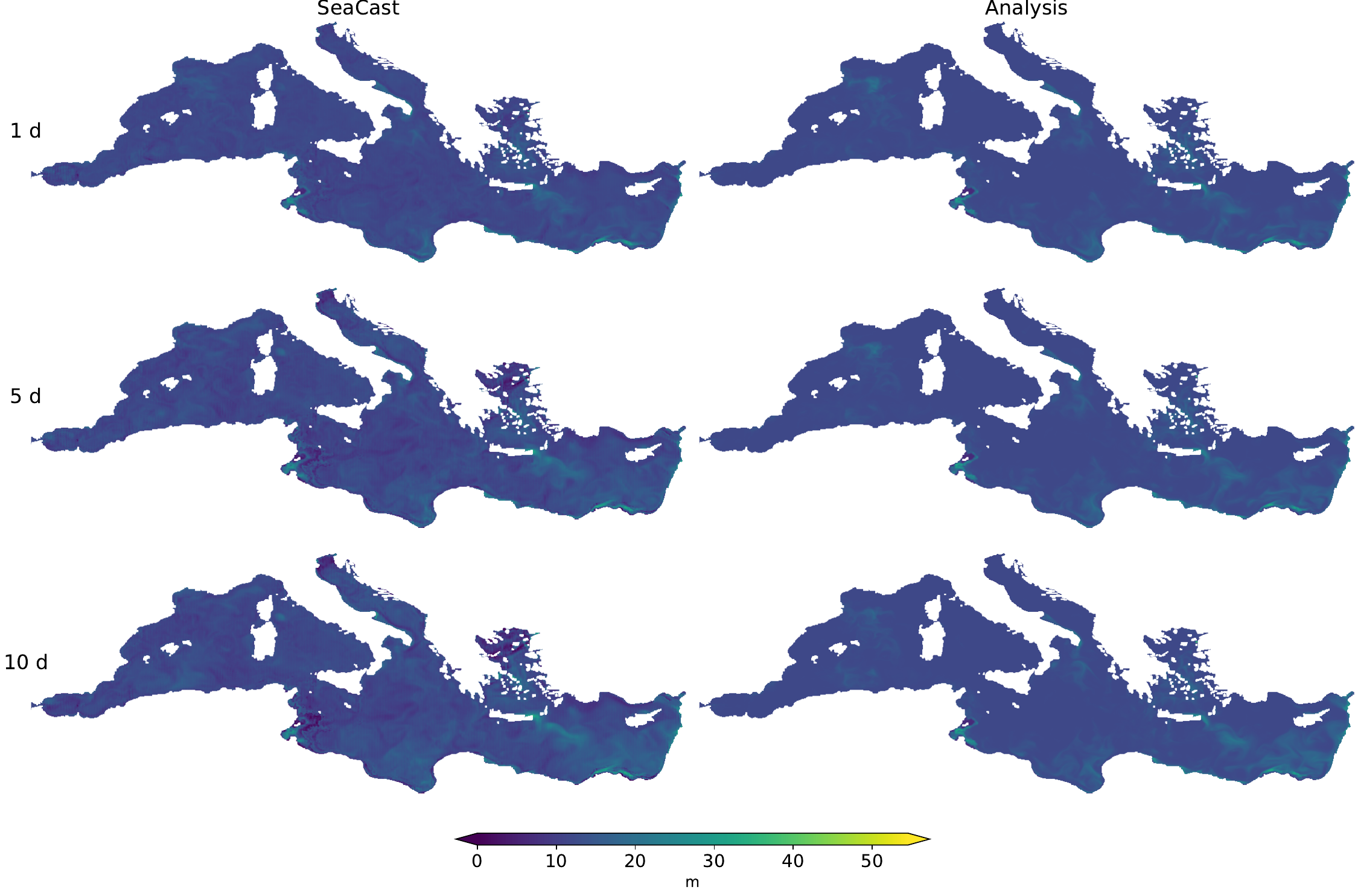}
    \caption{SeaCast-AIFS forecast for \texttt{mlotst} initialized on August 1st, 2024, vs. analysis.}
    \label{fig:rollout_mlotst}
\end{figure}

\begin{figure}[h]
    \centering
    \includegraphics[width=\textwidth]{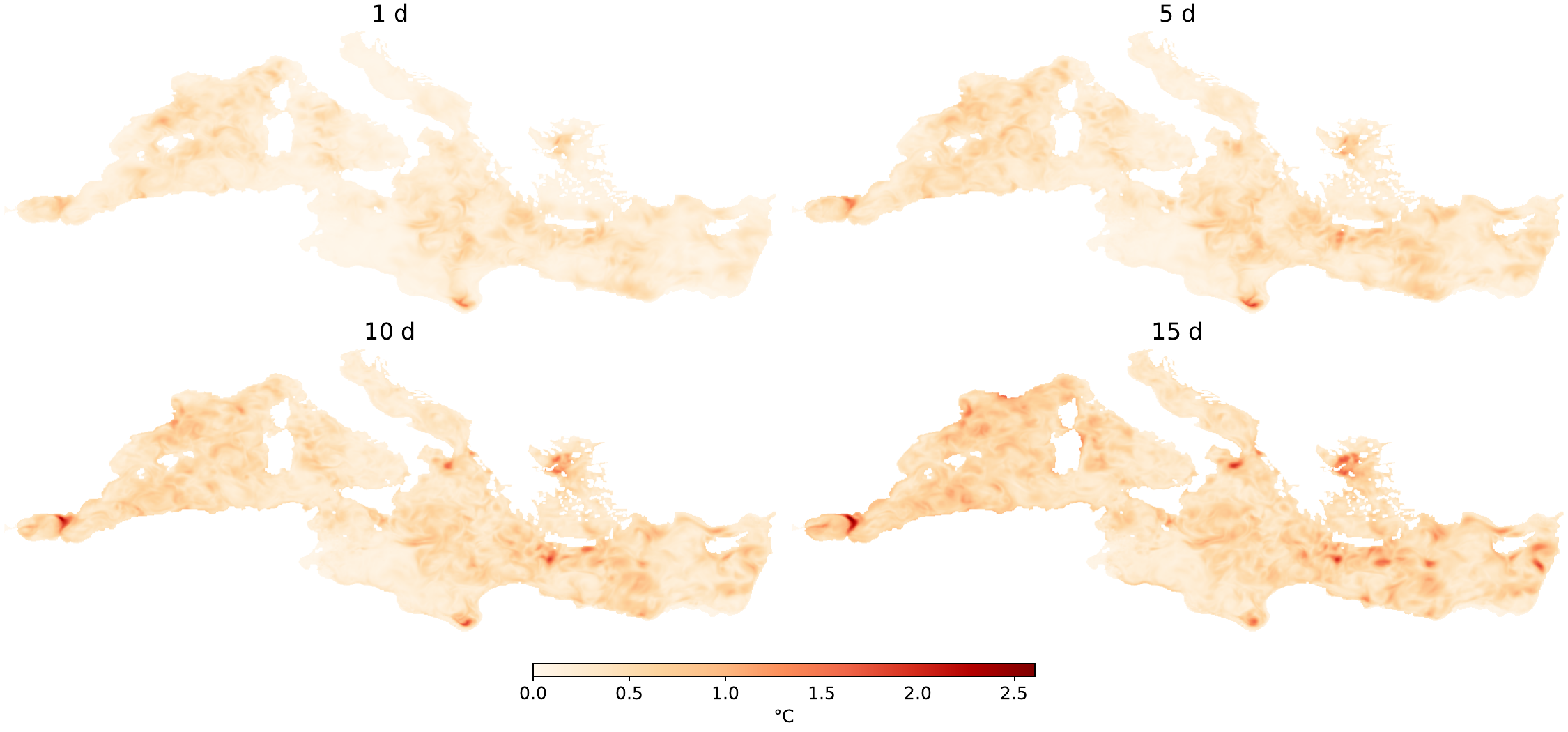}
    \caption{Spatial RMSE of SeaCast-AIFS \texttt{thetao} forecasts vs. analysis at different leads.}
    \label{fig:error_thetao}
\end{figure}

\begin{figure}[h]
    \centering
    \includegraphics[width=\textwidth]{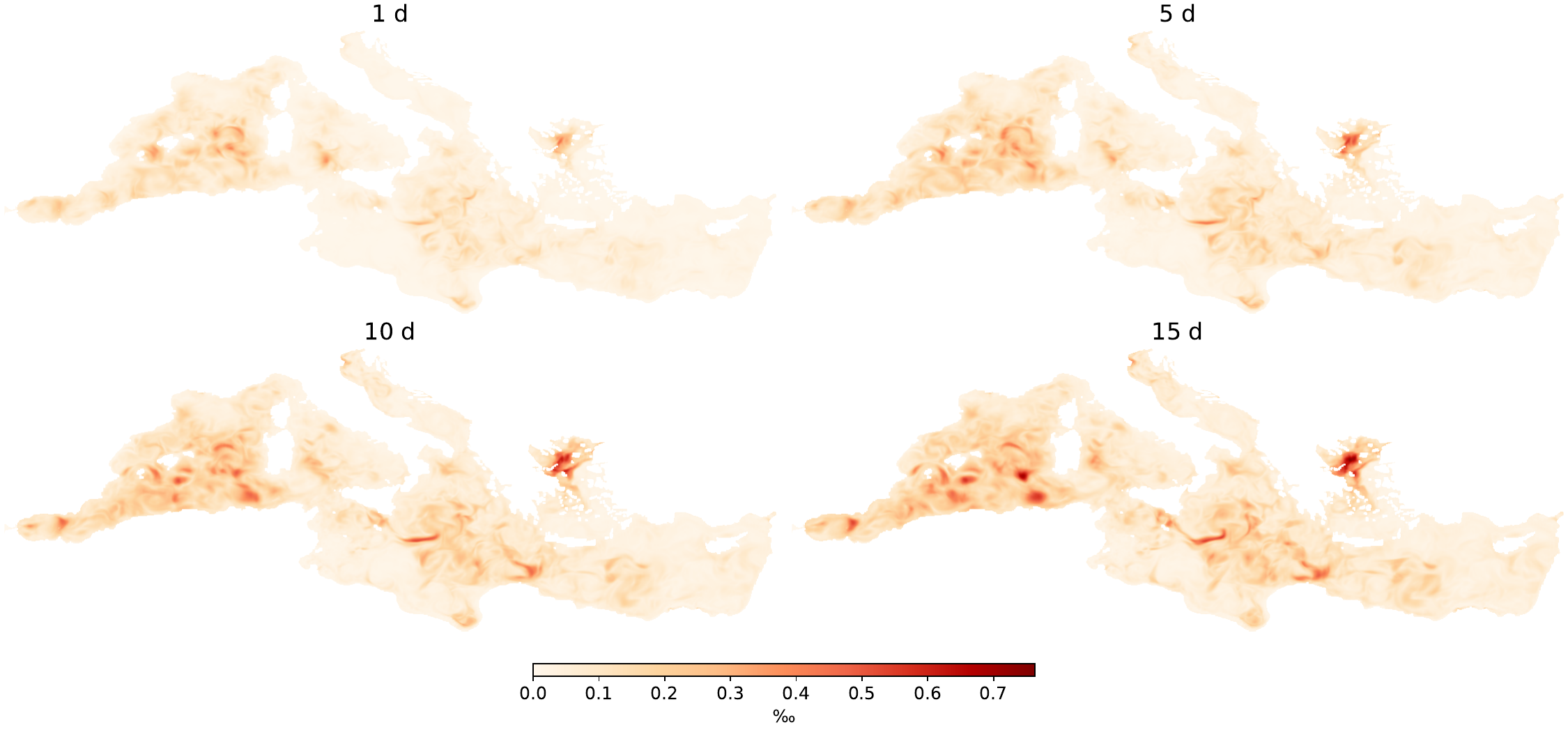}
    \caption{Spatial RMSE of SeaCast-AIFS \texttt{so} forecasts vs. analysis at different leads.}
    \label{fig:error_so}
\end{figure}

\begin{figure}[h]
    \centering
    \includegraphics[width=\textwidth]{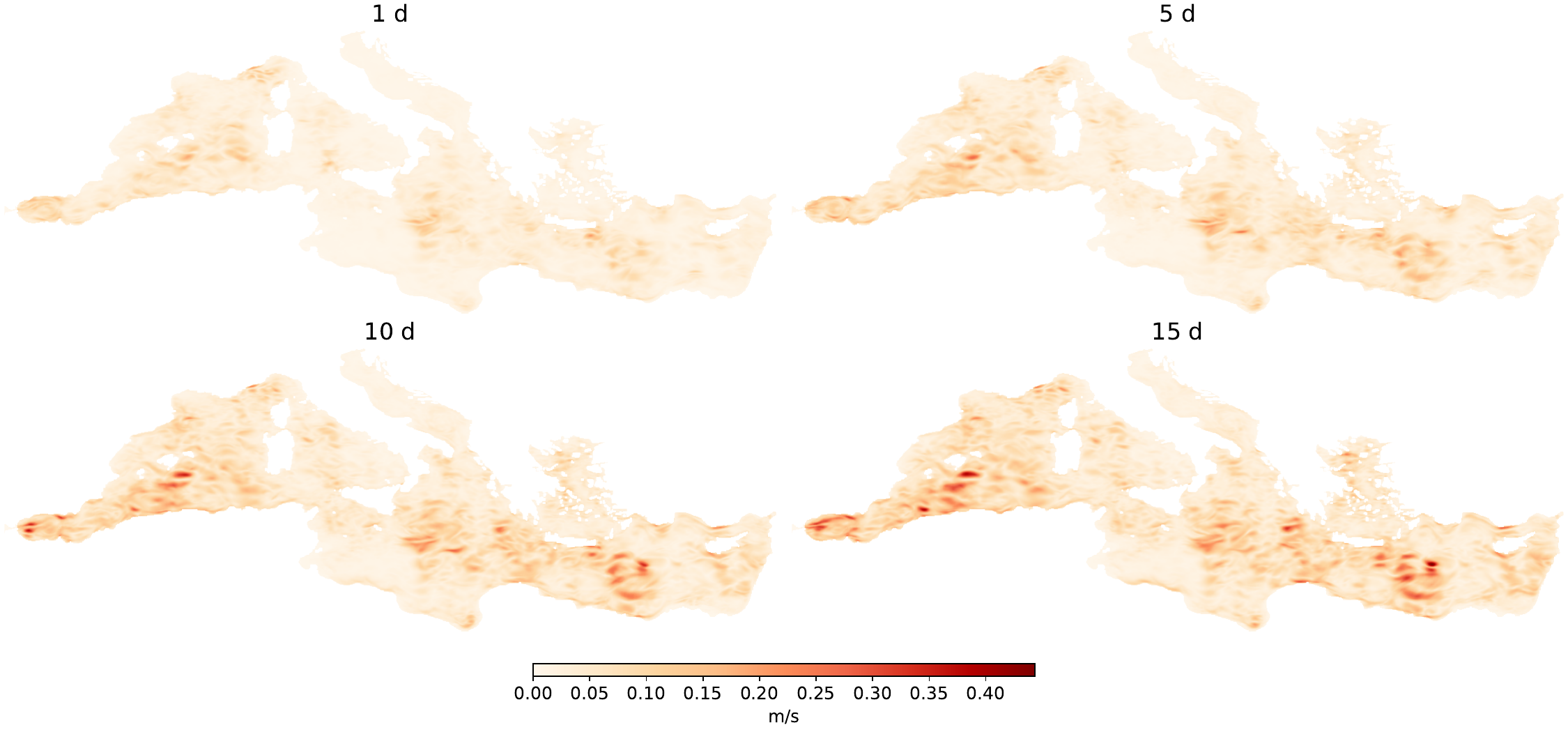}
    \caption{Spatial RMSE of SeaCast-AIFS \texttt{uo} forecasts vs. analysis at different leads.}
    \label{fig:error_uo}
\end{figure}

\begin{figure}[h]
    \centering
    \includegraphics[width=\textwidth]{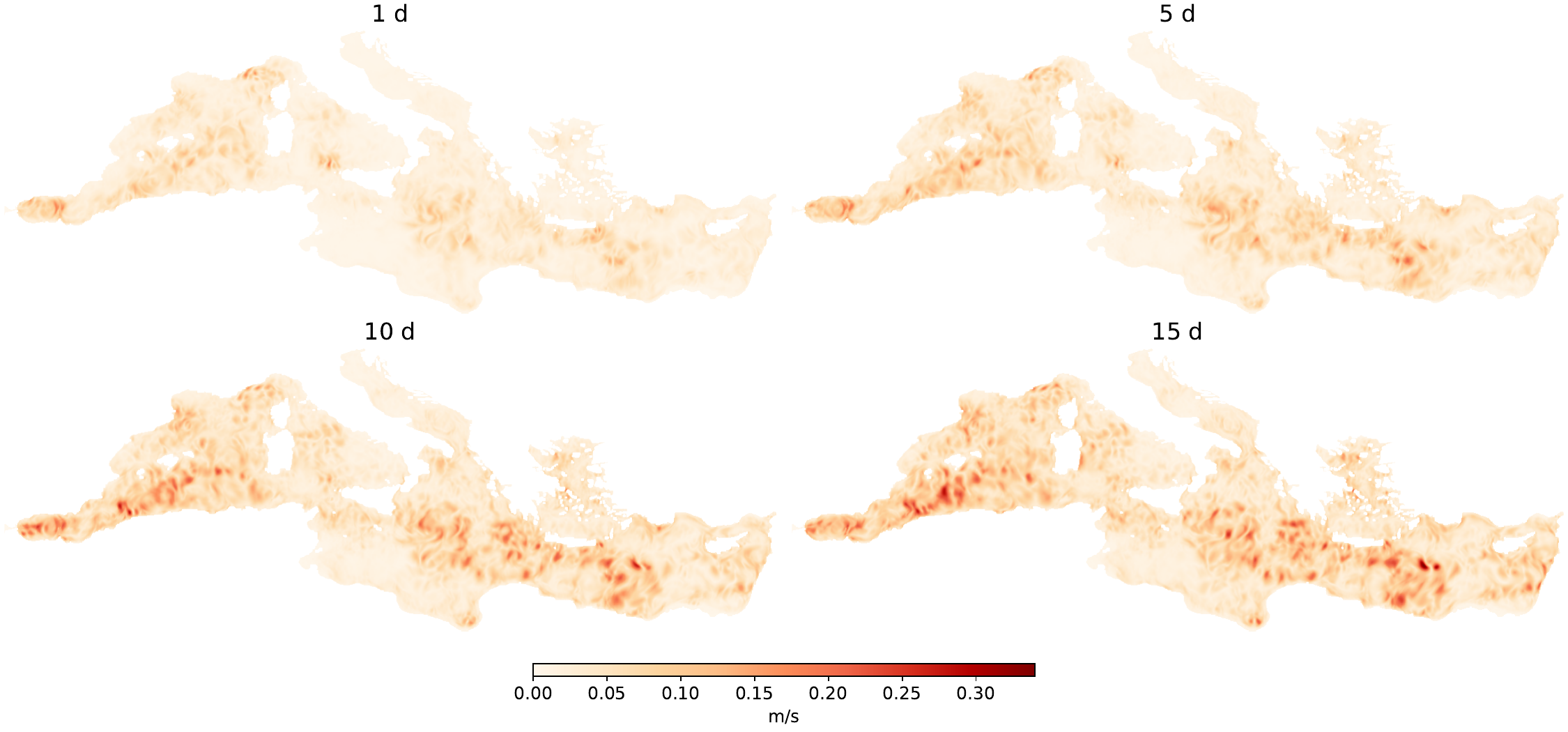}
    \caption{Spatial RMSE of SeaCast-AIFS \texttt{vo} forecasts vs. analysis at different leads.}
    \label{fig:error_vo}
\end{figure}

\begin{figure}[h]
    \centering
    \includegraphics[width=\textwidth]{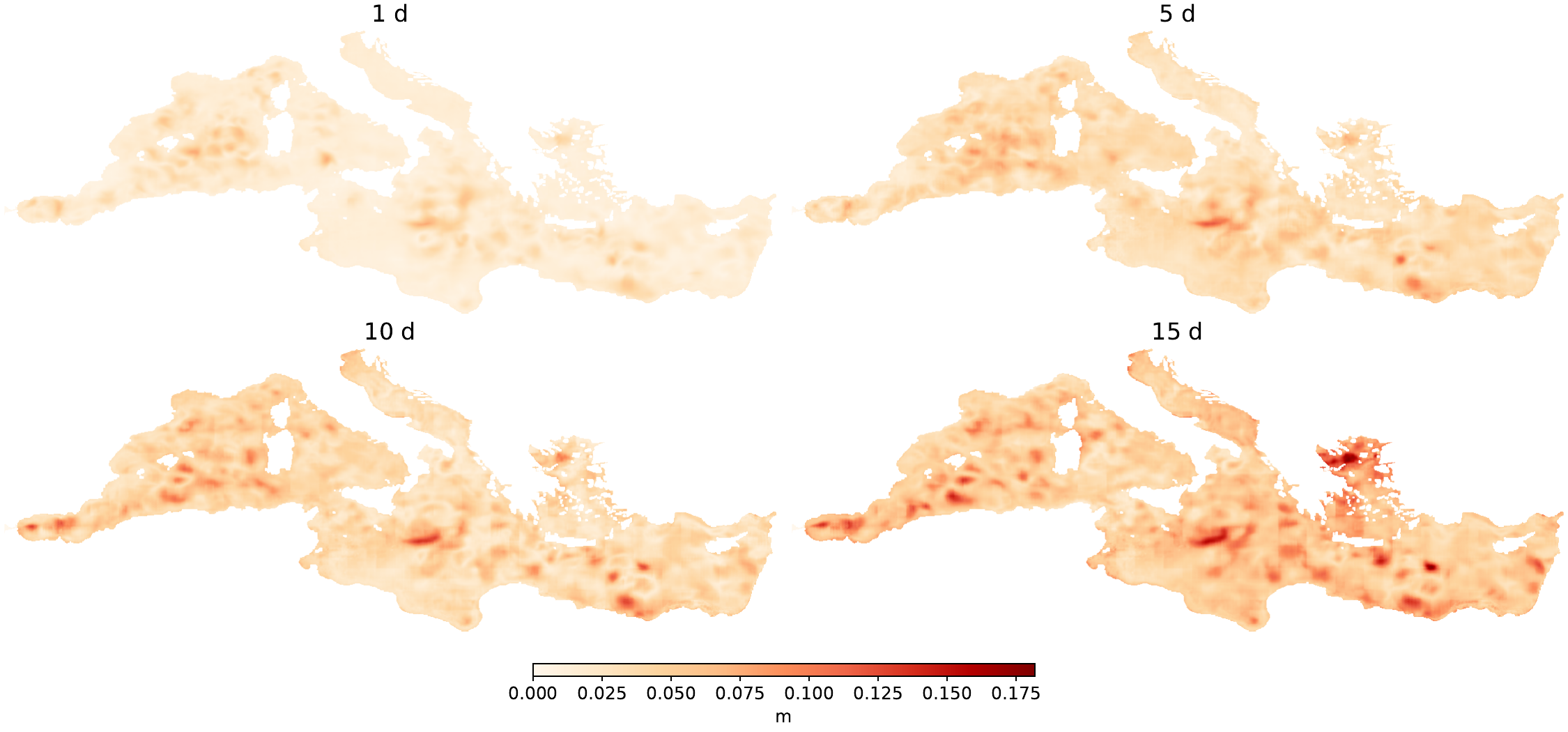}
    \caption{Spatial RMSE of SeaCast-AIFS \texttt{zos} forecasts vs. analysis at different leads.}
    \label{fig:error_zos}
\end{figure}

\begin{figure}[h]
    \centering
    \includegraphics[width=\textwidth]{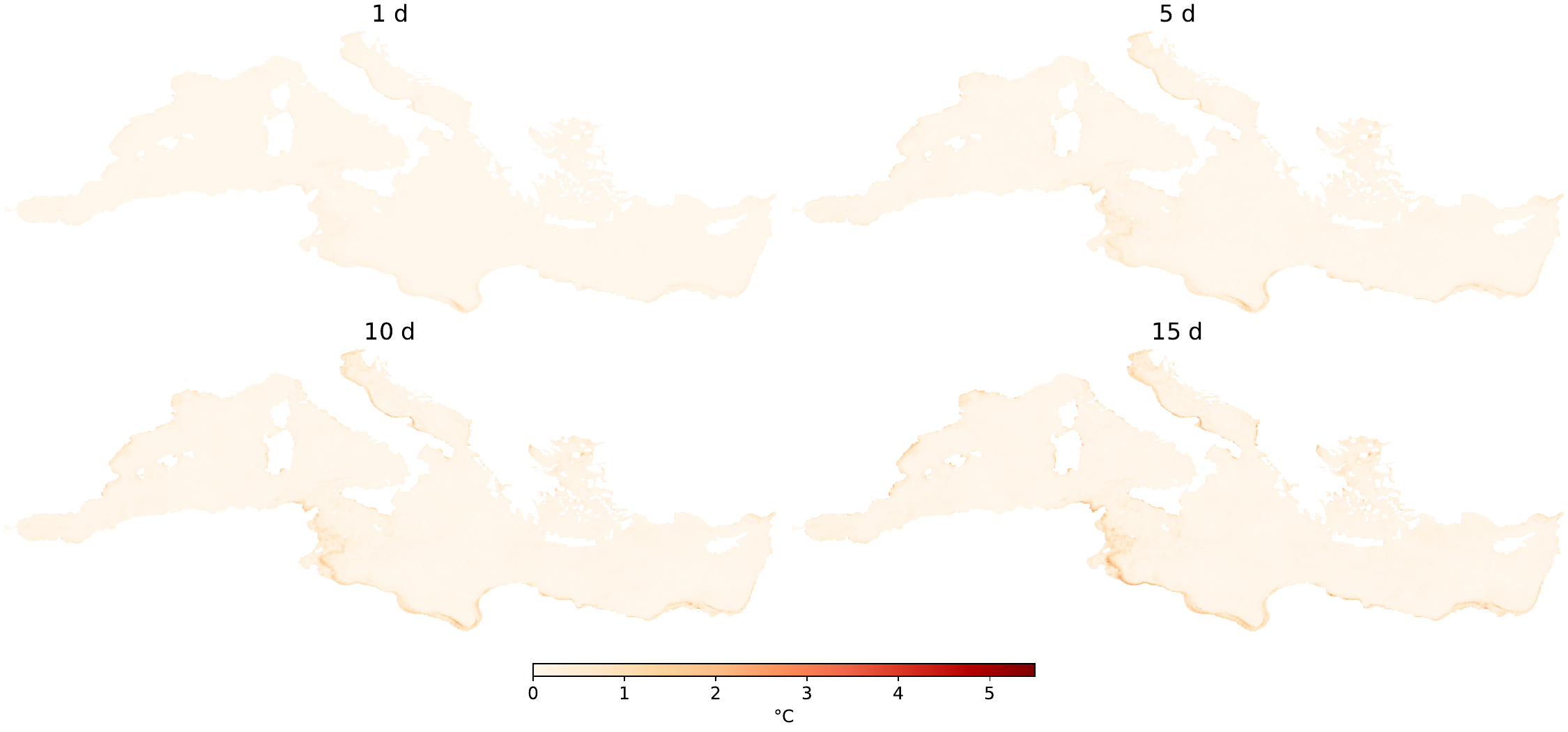}
    \caption{Spatial RMSE of SeaCast-AIFS \texttt{bottomT} forecasts vs. analysis at different leads.}
    \label{fig:error_bottomT}
\end{figure}

\begin{figure}[h]
    \centering
    \includegraphics[width=\textwidth]{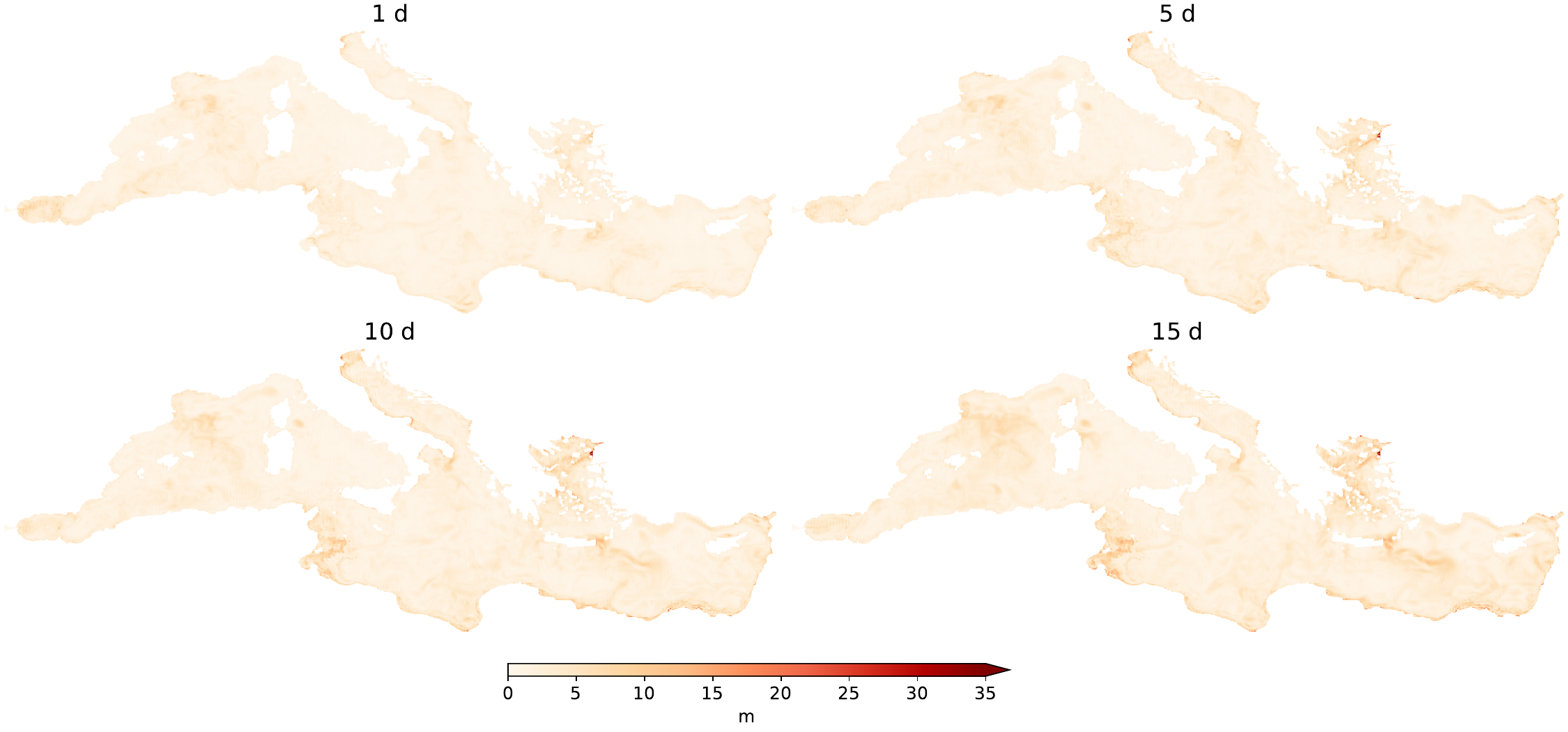}
    \caption{Spatial RMSE of SeaCast-AIFS \texttt{mlotst} forecasts vs. analysis at different leads.}
    \label{fig:error_mlotst}
\end{figure}

Figures~\ref{fig:rollout_thetao_30} to \ref{fig:rollout_mlotst} provide example visualizations of forecasts produced with SeaCast-AIFS, while Figures~\ref{fig:error_thetao} to \ref{fig:error_mlotst} display the corresponding spatial errors. The mixed layer depth (\texttt{mlotst}) is generally shallow during the summer period considered here due to the stratification of the water column, but there are some highly localized errors, particularly near the Dardanelles Strait. This may be because the reanalysis employs a closed boundary and does not model tides, whereas the analysis and forecast system features an open boundary there and includes tidal inputs. Overall, for spatial errors, we observe that open sea regions or areas with stronger currents tend to have larger errors over time. Since RMSE is a measure of error magnitude, higher values indicate greater absolute errors when the model's predictions deviate from the true values.

\bibliographyapp{references}
\bibliographystyleapp{unsrt}

\clearpage

\end{document}